\documentclass[12pt]{iopart}
\usepackage{dcolumn,graphicx,color,booktabs,microtype,afterpage}
\usepackage{float}
\usepackage{amssymb}
\usepackage{url}  
\usepackage[charter,greekuppercase=italicized]{mathdesign}
\usepackage{sidecap}
\usepackage[mathlines]{lineno}
\graphicspath{{./}{figure/}}
\newcommand{\tcr}[1]{\textcolor{black}{#1}}

\begin{document}

\title[Multigap SC in the Mo$_5$PB$_2$ boron-phosphorus compound]{\tcr{Multigap superconductivity in the Mo$_5$PB$_2$ boron-\\phosphorus compound}}


\author{T\ Shang$^1$, W\ Xie$^2$, D\ J\ Gawryluk$^1$, R\ Khasanov$^3$, J\ Z\ Zhao$^{4,5}$,
        M\ Medarde$^1$,  M\ Shi$^6$, H\ Q\ Yuan$^{2,7}$, E\ Pomjakushina$^1$, T~Shiroka$^{3,8}$}

\address{$^1$Laboratory for Multiscale Materials Experiments, Paul Scherrer Institut, Villigen CH-5232, Switzerland}
\address{$^2$Center for Correlated Matter and Department of Physics, Zhejiang University, Hangzhou, 310058, China}
\address{$^3$Laboratory for Muon-Spin Spectroscopy, Paul Scherrer Institut, CH-5232 Villigen PSI, Switzerland}
\address{$^4$Co-Innovation Center for New Energetic Materials, Southwest University of Science and Technology, Mianyang, 621010, People's Republic of China} 
\address{$^5$Research Laboratory for Quantum Materials, Singapore University of Technology and Design, Singapore 487372, Singapore} 
\address{$^6$Swiss Light Source, Paul Scherrer Institut, Villigen CH-5232, Switzerland}
\address{$^7$Collaborative Innovation Center of Advanced Microstructures, Nanjing, 210093, China}
\address{$^8$Laboratorium f\"ur Festk\"orperphysik, ETH Z\"urich, CH-8093 Z\"urich, Switzerland}
\ead{tian.shang@psi.ch}

\begin{indented}
\item
\today
\end{indented}

\begin{abstract}
The tetragonal Mo$_5$PB$_2$ compound was recently reported 
to show su\-per\-con\-duc\-ti\-vi\-ty with a critical temperature up to 9.2\,K. 
In search of evidence for multiple superconducting gaps in Mo$_5$PB$_2$, 
comprehensive measurements, including magnetic susceptibility, electrical 
resistivity, heat capacity, and muon-spin rotation and relaxation 
($\mu$SR) measurements were carried out. 
Data from both low-temperature superfluid density and  
electronic specific heat suggest a nodeless superconducting ground 
state in Mo$_5$PB$_2$. Two superconducting energy gaps 
$\Delta_0$ = 1.02\,meV (25\%) 
and 1.49\,meV (75\%) 
are required to describe the low-$T$ electronic specific-heat data. 
The multigap features are clearly evidenced by the field dependence of 
the electronic specific-heat coefficient and the Gaussian relaxation rate 
in the superconducting state (i.e., superfluid density), as well as by the temperature dependence of the upper critical field. By combining our 
extensive experimental results with numerical band-structure calculations, 
we provide compelling
evidence of multigap superconductivity in Mo$_5$PB$_2$.
\end{abstract}

%
\vspace{2pc}
\noindent{\it Keywords}: Multigap, superconductivity, $\mu$SR 

%
\submitto{\NJP}
%
\maketitle
%
%

\section{\label{sec:Introduction}Introduction}

The $T_5M_3$ family, where $T$ is a transition or rare-earth 
metal and $M$ a (post)-transition metal or a metalloid element, 
features three distinct structural symmetries: orthorhombic Yb$_5$Sb$_3$-type ($Pnma$, No.\ 62), tetragonal Cr$_5$B$_3$-type ($I4/mcm$, No.\ 140), and hexagonal Mn$_5$Si$_3$-type ($P6_3/mcm$, No.\ 193). The tetragonal Cr$_5$B$_3$-type structure is adopted by a broad range of binary and ternary compounds. Among these, the layered 
ternary compounds of transition metals with boron and silicon 
(or boron and phosphorus),  
with a $T_5$$X$B$_2$ stoichiometry ($X$ = P or Si), 
exhibit many 
interesting properties. For example, Co$_5$SiB$_2$ exhibits a 
paramagnetic ground state, found to persist down to 
liquid He temperature~\cite{Cristina2010}. On the other hand, 
when $T$ is occupied by other $3d$ metals, such as Mn or Fe,  
both $T_5$SiB$_2$ and $T_5$PB$_2$ are ferromagnets with high Curie temperatures. 
Therefore, currently they are being studied for room-temperature magnetocaloric applications or as rare-earth-free permanent magnets~\cite{Almeida2009,Xie2010,McGuire2015,Lamichhane2016}. Unlike these high-temperature ferromagnets, the 4$d$ and 5$d$ compounds Nb$_5$SiB$_2$, Mo$_5$SiB$_2$, and W$_5$SiB$_2$ are superconductors, with transition temperatures in the 5 to 8\,K range~\cite{Brauner2009,Machado2011,Fukuma2011,Fukuma2012}. 
Later on, the Cr$_5$B$_3$-type Ta$_5$GeB$_2$ boro-germanide could 
also be synthesized and shown to become a superconductor 
below $T_c \sim 3.8$\,K~\cite{Correa2016}. 

Very recently, a new member of the Cr$_5$B$_3$-type series,  
namely Mo$_5$PB$_2$, was synthesized and shown to exhibit superconductivity 
(SC) with a critical temperature $T_c = 9.2$\,K~\cite{McGuire2016}, the highest 
$T_c$ recorded in this family of compounds. 
According to electrical resistivity measurements under various applied 
magnetic fields, its upper critical field, 
$\mu_0H_{c2} \sim 1.7$\,T, seems
to be much higher 
than that of Mo$_5$SiB$_2$ (0.6\,T) or W$_5$SiB$_2$ (0.5\,T)~\cite{Machado2011,Fukuma2011}. In addition, over 
a wide temperature range, the temperature-dependent $H_{c2}(T)$ 
of Mo$_5$PB$_2$ 
seems inconsistent with the Ginzburg-Landau- or Werthamer-Helfand-Hohenberg models, implying multiple superconducting gaps in Mo$_5$PB$_2$~\cite{McGuire2016}. 
To date, a detailed analysis of the $H_{c2}(T)$ data is 
still missing. Yet, possible multigap features were already suggested by 
zero-field heat-capacity measurements and electronic band-structure 
calculations~\cite{McGuire2016}. Indeed, its zero-field specific-heat 
seems more consistent with a two-gap-  
rather than with a single-gap model, as confirmed also by 
the present work. First-principle calculations indicate Mo$_5$PB$_2$ to be a multiband metal, whose density of states (DOS) at the Fermi level is dominated by the Mo 4$d$-orbitals.

Although electronic band-structure calculations are available for Mo$_5$PB$_2$ and its superconductivity has been studied 
via macroscopic techniques (e.g., specific heat), the microscopic 
nature of its SC remains largely unexplored. 
In particular, the multigap feature of Mo$_5$PB$_2$ demands stronger evidence. 
To this aim, we performed
an extensive study of the superconducting properties of Mo$_5$PB$_2$ by means of electrical resistivity, magnetization, thermodynamic- and, in particular, by muon-spin rotation and relaxation ($\mu$SR) methods.  
We find that Mo$_5$PB$_2$ exhibits a fully-gapped superconducting state with 
preserved time-reversal symmetry.
Its multigap features are strongly evidenced by the field-dependent 
electronic specific-heat coefficient, as well as by the superconducting 
$\mu$SR relaxation, the latter being highly consistent with the temperature dependence of the upper critical field.     

\section{Methods\label{sec:details}}

Polycrystalline samples of Mo$_5$PB$_2$ were prepared by solid-state 
reaction methods, the procedures used to synthesize 
the material being reported in detail elsewhere~\cite{McGuire2016}. 
Room-temperature x-ray powder diffraction (XRD) measurements 
were used to check the quality of the Mo$_5$PB$_2$ samples, by 
employing a Bruker D8 diffractometer with Cu K$\alpha$ radiation.
The magnetic susceptibility, electrical resistivity, and heat-capacity measurements were performed on a 7-T Quantum Design magnetic 
property measurement system (MPMS-7) and a 14-T physical property 
measurement system (PPMS-14) equipped with a $^3$He  cryostat. 

The bulk $\mu$SR measurements were carried out at
the general-purpose (GPS) and the multipurpose (Dolly) surface-muon
spectrometers at the Swiss muon source of Paul Scherrer
Institut, Villigen, Switzerland~\cite{Amato2017}. 
For the low-temperature measurements on Dolly (down to $\sim$0.3\,K), the 
samples were mounted on a thin copper foil (ca.\ $\sim$30\,$\mu$m thick) 
using diluted GE varnish. 
Transverse-field (TF) $\mu$SR measurements were carried out to 
investigate the superconducting properties (mostly the gap symmetry) of 
Mo$_5$PB$_2$. 
To track the additional field-distribution broadening due to the flux-line-lattice (FLL) in the mixed superconducting state, 
we followed a field-cooling (FC) protocol, where the magnetic field is 
applied in the normal state, before cooling the sample down to base temperature. 
Afterwards, the TF-$\mu$SR spectra were collected at various temperatures upon warming.
The $\mu$SR data were analyzed by means of the \texttt{musrfit} software package~\cite{Suter2012}.

The electronic band structure of Mo$_5$PB$_2$ was
calculated via the 
density functional theory, within the generalized gradient approximation 
(GGA) of Perdew-Burke-Ernzerhof (PBE) realization~\cite{Perdew1996iq}, 
as implemented in the Vienna ab-initio Simulation Package (VASP)~\cite{Kresse1996kl,Kresse1996vk}. The projector augmented wave (PAW) pseudopotentials were adopted for the calculation~\cite{Kresse1999wc,Blochl1994zz}. Electrons belonging to the outer atomic configuration 
were treated as valence electrons, here corresponding to 6 electrons in Mo ($4d^55s^1$), 5 electrons in P ($3s^23p^3s$), and 3 electrons in B ($2s^22p^1$). The kinetic energy cutoff was fixed to 500\,eV. The lattice parameters and the atomic positions 
experimentally determined from Rietveld refinements were chosen 
for the calculations. For the self-consistent calculation, the Brillouin 
zone integration was performed on a $\Gamma$-centered mesh of 
$10 \times 10 \times 10$ $k$-points.

\section{\label{sec:results} Results and discussion}
\subsection{\label{ssec:structure} Crystal structure and phase purity}
%
\begin{figure}[th]
	\centering
	\includegraphics[width=0.6\textwidth,angle=0]{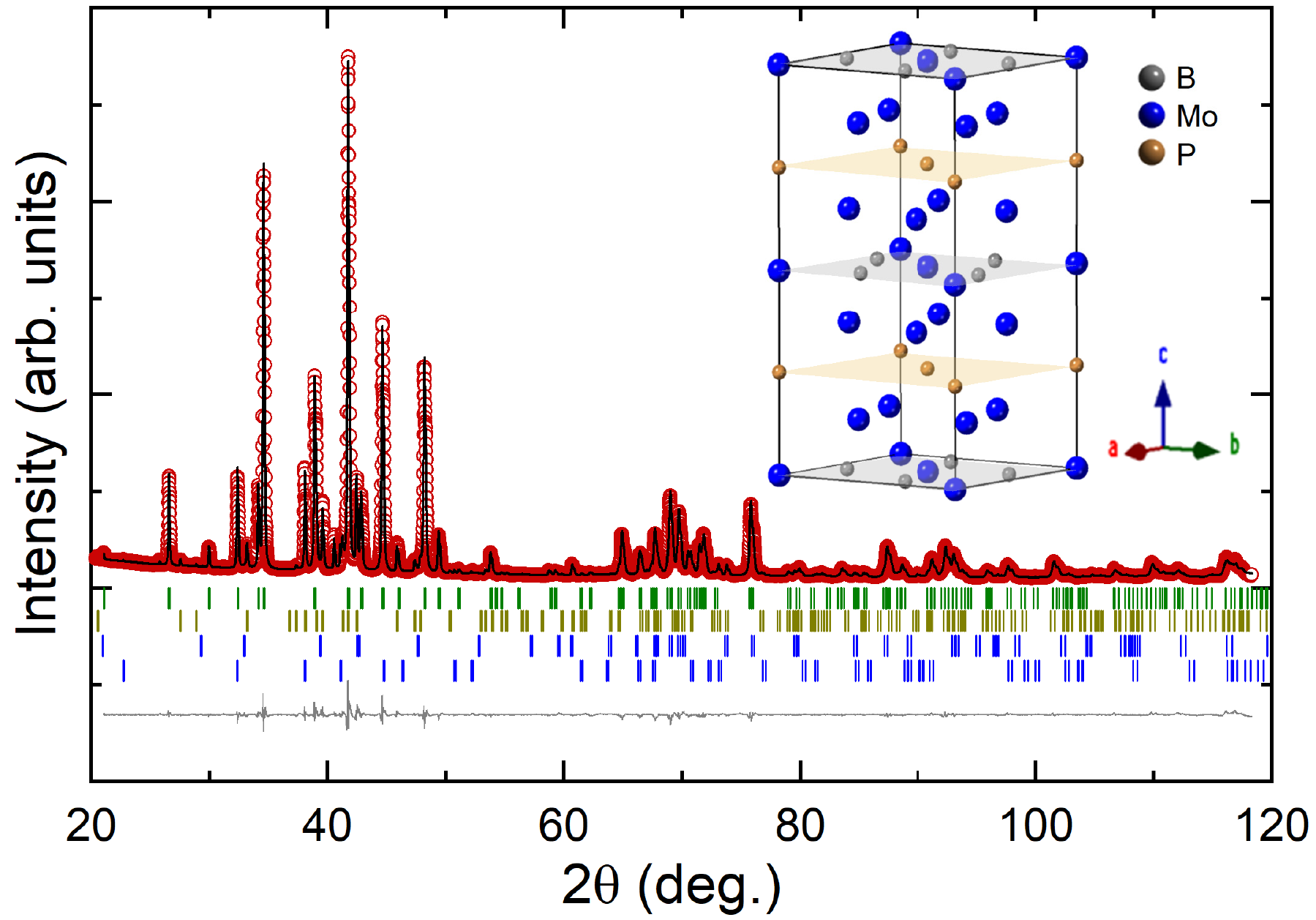}
	\vspace{-2ex}%
	\caption{\label{fig:XRD}Room-temperature x-ray powder diffraction pattern and
		Rietveld refinement for Mo$_5$PB$_2$. The red circles and the solid
		black line represent the experimental pattern and the Rietveld refinement
		profile, respectively. The gray line at the bottom shows
		the residuals, i.e., the difference between calculated and experimental
		data. The vertical bars mark the calculated Bragg-peak positions
		for Mo$_5$PB$_2$ (green), Mo$_3$P (yellow), and MoB/Mo$_2$B (blue). The crystal
		structure (unit cell) is shown in the inset.}
\end{figure}
%
The crystal structure and the purity of Mo$_5$PB$_2$ polycrystalline samples 
were checked via powder XRD at room temperature. Figure~\ref{fig:XRD} shows 
a refinement of the XRD pattern, performed by means of the FullProf Rietveld-analysis suite~\cite{Carvajal1993}. 
The refinement confirms that Mo$_5$PB$_2$ crystallizes in the tetragonal Cr$_5$B$_3$-type structure, also known as T$_2$-phase. 
The refined lattice parameters, $a = b = 5.97105(5)$\,\AA\ and $ c = 11.07008(11)$\,\AA, 
are in good agreement with the results reported in the literature~\cite{McGuire2016}.  
Similar to previous work, also our data (see figure~\ref{fig:XRD}) 
indicate that, besides the main Mo$_5$PB$_2$ phase (80\%), 
there are also extra reflections belonging to minor foreign phases: 
Mo$_3$P (16\%) and MoB/Mo$_2$B (4\%). 
Once formed, due to their very high melting temperature (above 
2000$^\circ$C), such extraneous phases are very stable and almost 
impossible to remove, 
even after multiple additional annealings. 
These minor phases, too, are superconductors, with critical temperatures 
below 5.5\,K~\cite{Ziegler1953,Matthais1952,Shang2019}. 
Nevertheless, upon investigating 
the Mo$_5$PB$_2$ samples, no superconducting 
signal from the MoB or Mo$_2$B phases could be identified. Therefore, 
they do not influence the determination of the superconducting 
parameters of Mo$_5$PB$_2$. As for Mo$_3$P, its contribution was 
properly subtracted when analyzing the zero-field specific-heat 
data (see details below and also in Ref.~\cite{McGuire2016}). 
The refined Mo$_5$PB$_2$ crystal structure, shown in the inset, 
comprises three different layers (MoB, Mo, and P), stacked 
alternatively along the $c$-axis and resembling a quasi-two 
dimensional 
structure. Clearly, in the unit cell there are two distinct 
crystallographic sites for the Mo atoms and a single site for the P or B atoms.

\subsection{\label{ssec:rho}Electrical resistivity}
%
%
\begin{figure}[th]
	\centering
	\includegraphics[width=0.6\textwidth,angle=0]{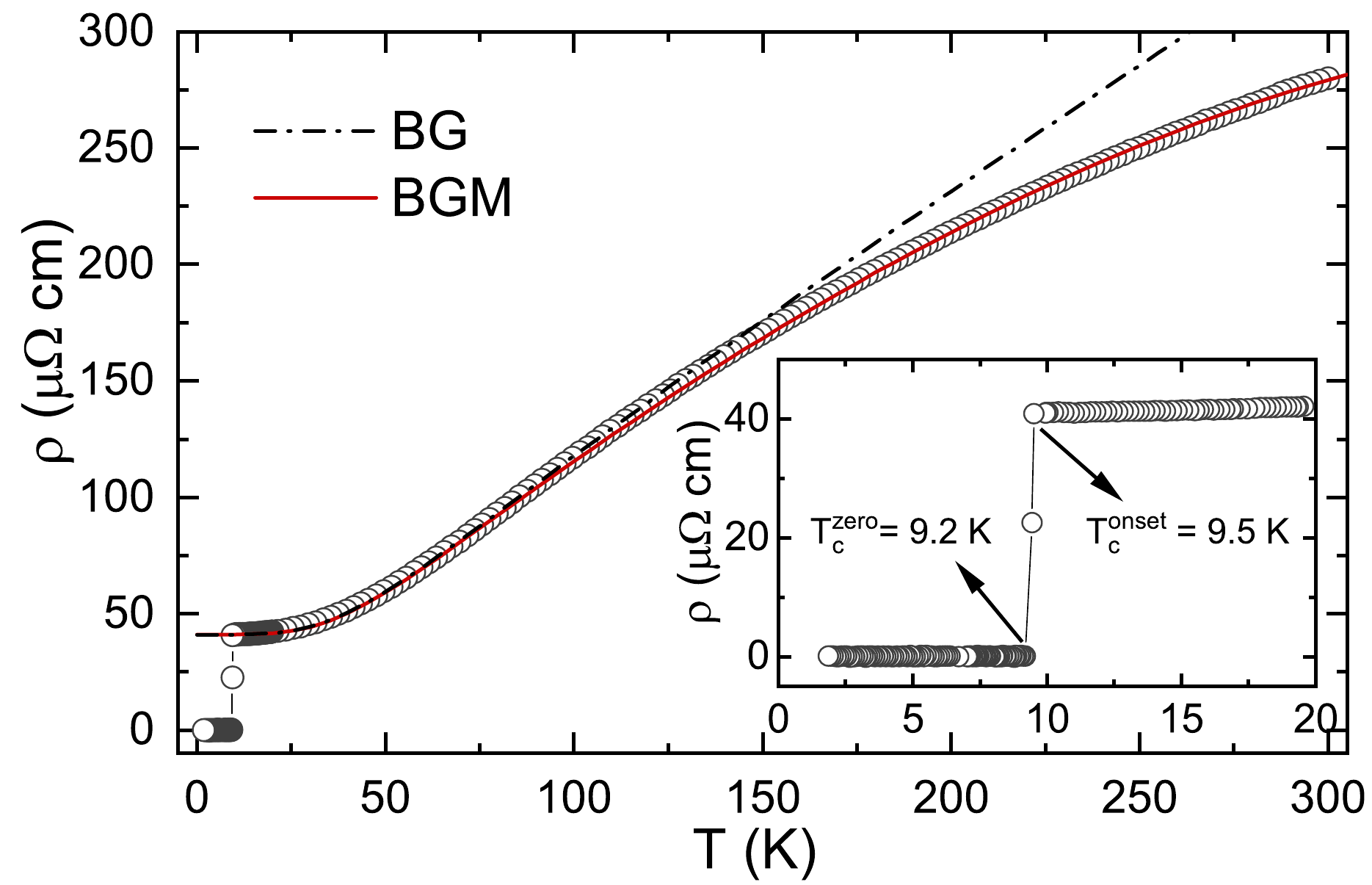}
	\vspace{-2ex}%
	\caption{\label{fig:Rho}Temperature dependence of the electrical 
		resistivity of Mo$_5$PB$_2$ collected in zero field up to room temperature. The black dash-dotted and red solid lines through the data are fits to the 
		Bloch-Gr\"{u}neisen formula with (BGM) and without (BG) Mott correction, respectively.
		The inset shows a closeup of the low-temperature region, highlighting the superconducting transition. }
\end{figure}
%
%
The temperature dependence of the electrical resistivity $\rho(T)$, 
collected in zero magnetic field from 300 down to 2\,K, reveals the 
metallic character of Mo$_5$PB$_2$ (see figure~\ref{fig:Rho}). 
The electrical resistivity in the low-$T$ region is shown in the 
inset. Here, the superconducting transition, with $T_c^\mathrm{onset} = 9.5$\,K and $T_c^\mathrm{zero} = 9.2$\,K, is clearly visible and the data are 
consistent with previous work~\cite{McGuire2016}.
The normal-state electrical resistivity is well modeled by the 
Bloch-Gr\"{u}neisen-Mott (BGM) formula 
$\rho(T) = \rho_0 + 4A (T/\Theta_\mathrm{D}^\mathrm{R})^5\int_0^{\Theta_\mathrm{D}^\mathrm{R}/T}\!\!\frac{z^5\mathrm{d}z}{(e^z-1)(1-e^{-z})}
-\alpha T$~\cite{Bloch1930,Blatt1968}. Here, $\rho_0$ represents the 
residual resistivity, while the second term describes the electron-phonon 
scattering, with $\Theta_\mathrm{D}^\mathrm{R}$ being the characteristic 
Debye temperature and $A$ a coupling constant.
The third term represents a contribution from the $s$-$d$ interband scattering, $\alpha$ being the Mott coefficient~\cite{Mott1958,Mott1964}. 
As shown in figure~\ref{fig:Rho}, the Mott correction is 
clearly required. Indeed, the black dash-dotted line, a fit to the BG 
formula without the Mott term, deviates significantly from the 
experimental data above 150\,K. 
The fit to BGM (red solid line) results in 
$\rho_0 = 41.1(2)$\,$\mu\mathrm{\Omega}$cm,
$A = 250(8)$\,$\mu\mathrm{\Omega}$cm, 
$\Theta_\mathrm{D}^\mathrm{R} =236(5)$\,K, and 
$\alpha = 2.2(1)$\,$\times$10$^{-6}$\,$\mu\mathrm{\Omega}$cmK$^{-3}$. 
A similar $\alpha$ value was also found in Mo$_3$P 
(3.4\,$\mu\mathrm{\Omega}$\,cm\,K$^{-3}$)~\cite{Shang2019},
indicating that, most likely, the $s$-$d$ scattering 
is due to Mo $d$- and to P $s$-electrons.

\subsection{\label{ssec:Cp_zero}Heat capacity} 

\begin{figure}[!thp]
	\centering
	\includegraphics[width=0.6\textwidth,angle=0]{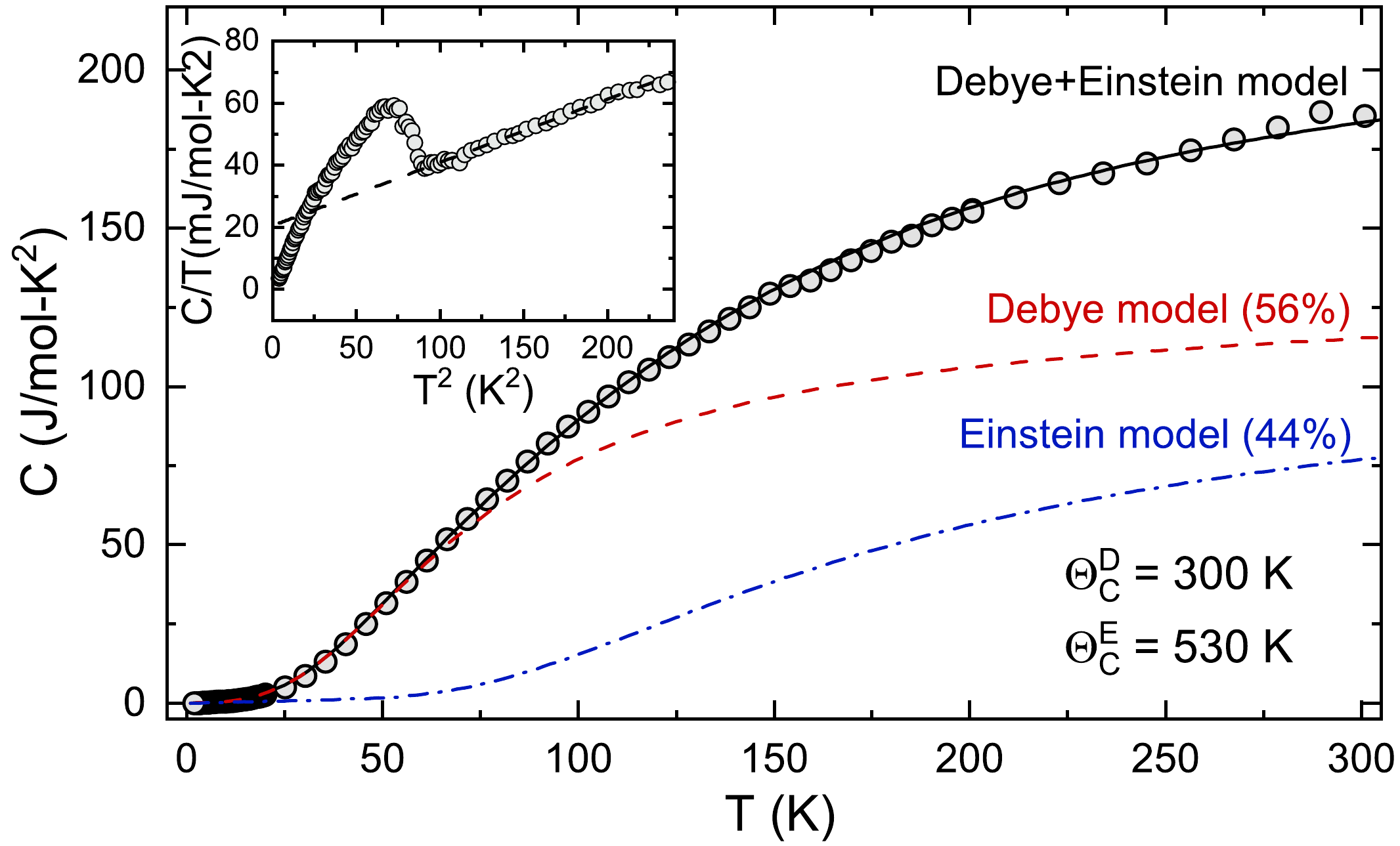}
	\vspace{-2ex}%
	\caption{\label{fig:Cp1}Temperature dependence of the Mo$_5$PB$_2$ 
		heat capacity, measured in zero field from 2 to 300\,K. 
		The solid line represents a fit to a combined Debye- and Einstein 
		model, with the dashed- and dash-dotted lines referring to the 
		two components.
		Inset: specific heat $C/T$ vs $T^2$ in the low-$T$ regime; the dashed-line is a fit to $C/T = \gamma_n + \beta T^2 + \delta T^4$,
		where $\gamma_\mathrm{n}$ is the electronic specific-heat coefficient, 
		while the two other terms account for the phonon contribution to the 
		specific heat. The determined values are $\gamma_\mathrm{n} = 22.3(2)$\,mJ/mol-K$^2$, 
		$\beta = 0.17(3)$\,mJ/mol-K$^4$ and $\delta = 1.3(9)\times 10^{-4}$\,mJ/mol-K$^6$.} 
\end{figure}
%
The Debye temperature can also be estimated from the heat capacity measurements. 
As shown in figure~\ref{fig:Cp1}, a pure Debye model cannot fit 
the $C(T)$ data properly. However, when combined with an 
Einstein model, it reproduces the $C(T)$ data fairly accurately. 
In this case, the solid line 
is a fit to the De\-bye- and Einstein model $C(T) = \gamma_\mathrm{n} T + n [\tcr{v}C_\mathrm{D}(T) + \tcr{(1-v)}C_\mathrm{E}(T)]$, with relative weights \tcr{$v$} and \tcr{$(1-v)$}. 
Here, $n = 8$ is the number of atoms per formula-unit in Mo$_5$PB$_2$. 
The first term represents the electronic specific heat, which 
can be extracted from the low-$T$ data (see inset in 
figure~\ref{fig:Cp1}).
The second and the third terms represent the acoustic- and optical 
phonon-mode contributions, described by the De\-bye- 
$C_\mathrm{D}(T) = 9R(T/\Theta_\mathrm{D}^\mathrm{C})^3\int_0^{\Theta_\mathrm{D}^\mathrm{C}/T}\!\!\frac{z^{4}e^z\mathrm{d}z}{(e^z-1)^2}$ 
and Einstein model $C_\mathrm{E}(T) = 3R(\Theta_\mathrm{E}^\mathrm{C}/T)^2\frac{\mathrm{exp}(\Theta_\mathrm{E}^\mathrm{C}/T)}{[\mathrm{exp}(\Theta_\mathrm{E}^\mathrm{C}/T)-1]^2}$,
respectively~\cite{Tari2003}.
Here $R = 8.314$\,J/mol-K is the molar gas constant, while 
$\Theta_\mathrm{D}^\mathrm{C}$ and $\Theta_\mathrm{E}^\mathrm{C}$ 
are the Debye and Einstein temperatures. The solid line in 
figure~\ref{fig:Cp1} represents the best fit, corresponding to
$\Theta_\mathrm{D}^\mathrm{C} = 300(5)$\,K, 
$\Theta_\mathrm{E}^\mathrm{C} = 530(5)$\,K, and $\tcr{v} = 0.56$. 
The obtained Debye temperature is slightly higher than that derived 
from electrical resistivity data (see figure~\ref{fig:Rho}). 
In fact, unlike electrical transport, heat capacity 
reflects better the bulk properties and, therefore, is more 
susceptible to extrinsic phases. 
In our case, the higher 
Debye temperature determined from 
heat-capacity measurements is most likely 
related to the MoB or Mo$_2$B phases, since the light B atoms usually exhibit rather high phonon frequencies, corresponding to 
large Debye temperatures (e.g., $\Theta_\mathrm{D} \sim$400\,K for MoB)~\cite{Rajpoot2018}.  

\subsection{\label{ssec:sus}Magnetization}
%
%
\begin{figure}[th]
	\centering
	\includegraphics[width=0.6\textwidth,angle=0]{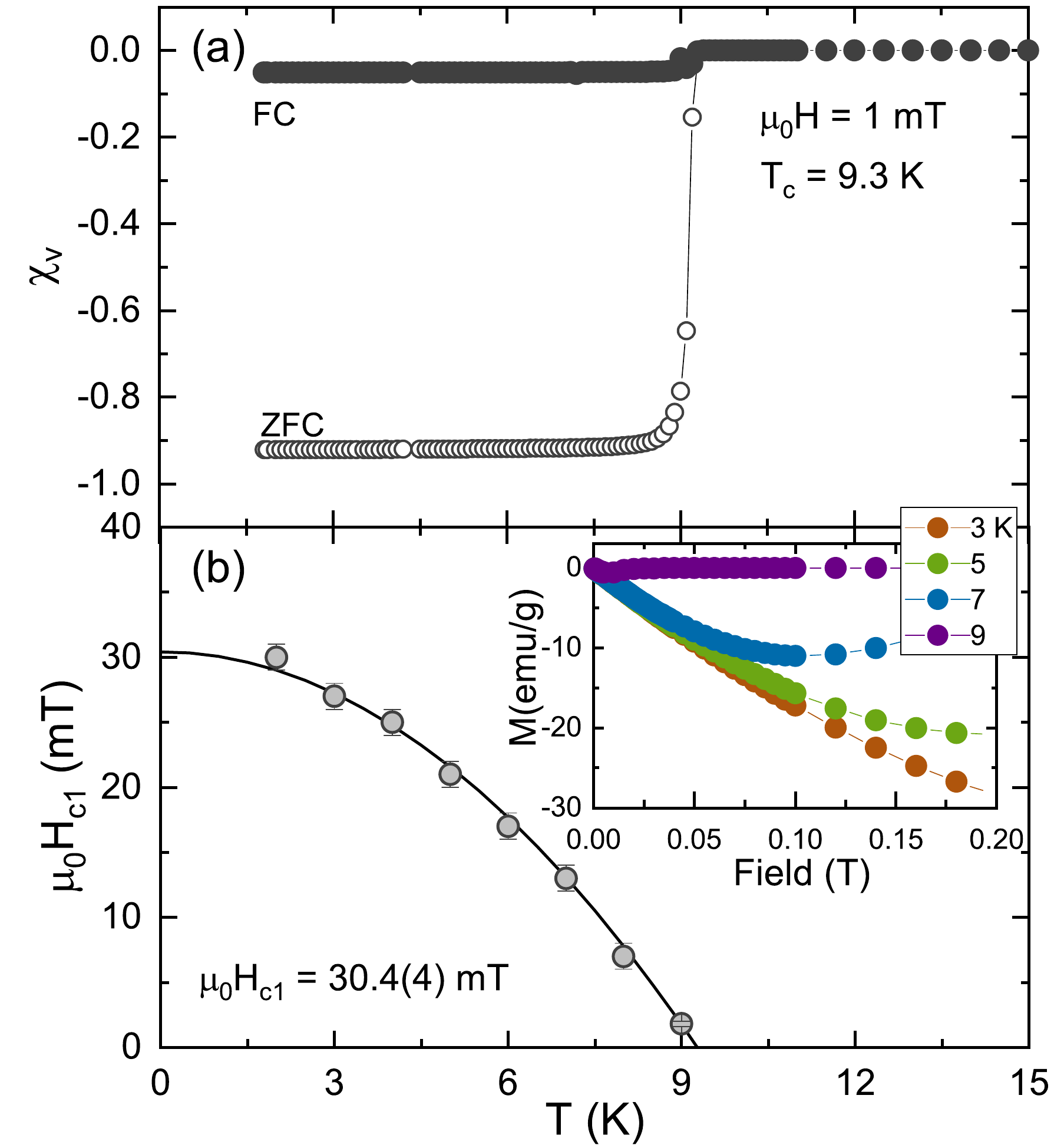}
	\vspace{-2ex}%
	\caption{\label{fig:Chi}(a) Temperature-dependent magnetic susceptibility of Mo$_5$PB$_2$, measured in an applied field of 1\,mT using both ZFC- and FC protocols. 
		(b) Estimated lower critical field $\mu_{0}H_{c1}$ vs temperature. 
		The solid line is a fit to 
		$\mu_{0}H_{c1}(T) =\mu_{0}H_{c1}(0)[1-(T/T_{c})^2]$. 
		The inset shows representative field-dependent magnetization 
		curves $M(H)$ recorded 
		at various temperatures up to $T_c$. For each temperature, the 
		lower critical field $\mu_{0}H_{c1}$ was determined as the 
		magnetic field where the diamagnetic response deviates from 
		the linear relation vs the magnetic field.}
\end{figure}

The superconductivity of Mo$_5$PB$_2$ was also evidenced by magnetization measurements.
The temperature-dependent magnetic susceptibility $\chi(T)$ measured in a field of 1\,mT using both field-cooled 
and zero-field-cooled (ZFC) protocols, 
is shown in figure~\ref{fig:Chi}(a). A sharp diamagnetic transition at 
$T_c$ = 9.3\,K indicates the onset of superconductivity in Mo$_5$PB$_2$, 
in agreement with the values determined from electrical resistivity and 
heat capacity. 
The well separated ZFC- and FC-susceptibility curves imply a strong 
flux-pinning effect in Mo$_5$PB$_2$.  
By assuming a cuboid 
(or, in general, an ellipsoid) sample shape 
with $a/b\sim$1 and $c/a\sim$0.5, the estimated demagnetization factor 
is $\sim$0.5, with the field applied along the $c$-direction~\cite{Amikam1998,Osborn1945}. After accounting for the demagnetization factor, 
the superconducting shielding fraction of  Mo$_5$PB$_2$ is about 92\%.  
To determine the lower critical field $\mu_{0}H_{c1}$ of Mo$_5$PB$_2$, 
essential for performing $\mu$SR measurements on type-II superconductors, 
the field-dependent magnetization $M(H)$ was measured at various temperatures up to $T_c$. 
Some representative $M(H)$ curves, recorded using a ZFC-protocol, are shown in the inset of figure~\ref{fig:Chi}(b). 
The estimated $\mu_{0}H_{c1}$ values vs temperature are summarized 
in the main panel, where the zero-temperature lower critical field
$\mu_{0}H_{c1}(0)$ = 30.4(4)\,mT is also determined. This is 
highly consistent 
with 30.8\,mT, the value calculated from the magnetic penetration
depth $\lambda_0$ (see below).

\subsection{\label{ssec:TF_muSR} TF-$\mu$SR and superconducting order parameter}
%
%
\begin{figure}[!thp]
	\centering
	\includegraphics[width=0.48\textwidth,angle= 0]{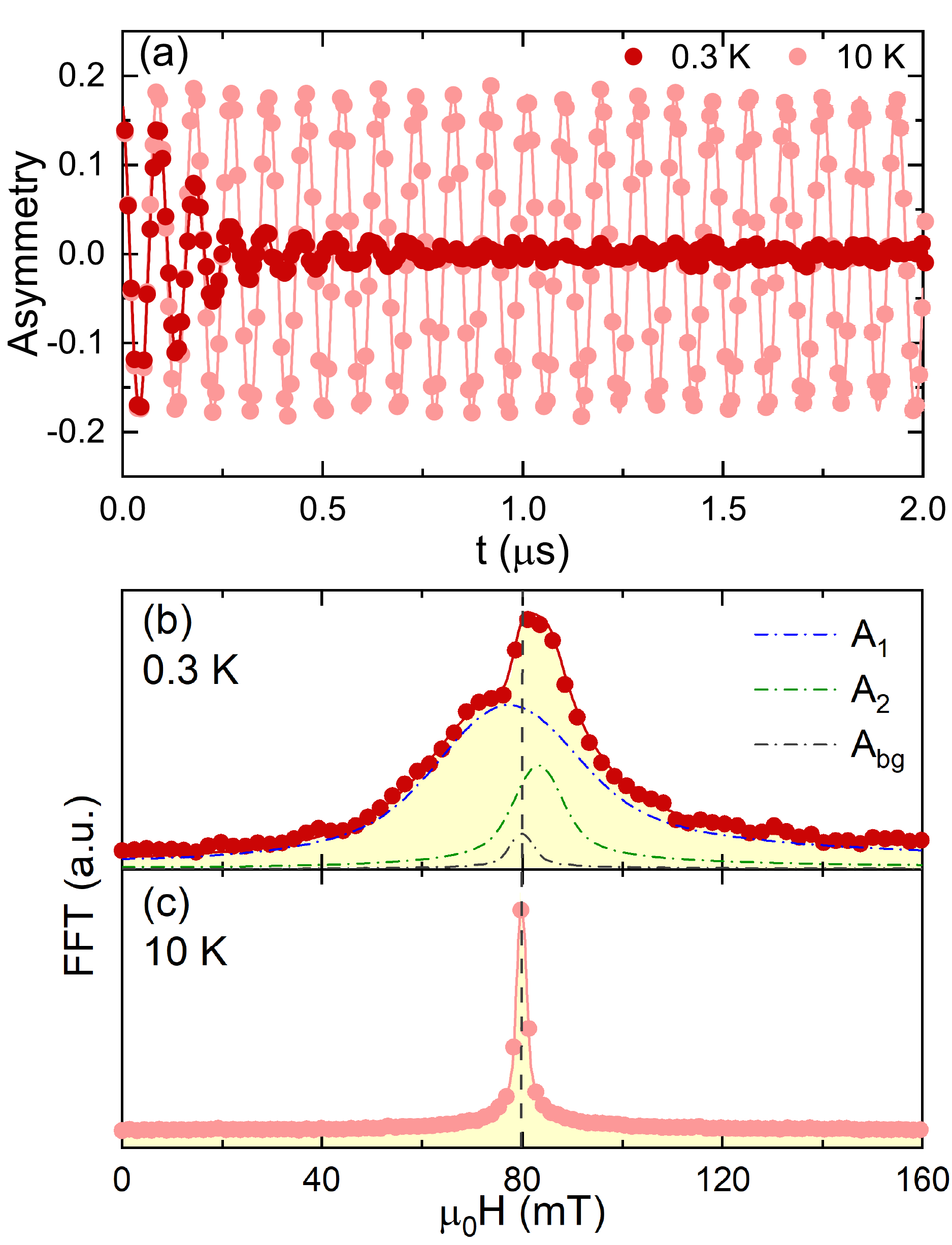}
	\caption{\label{fig:TF-muSR_T}(a) TF-$\mu$SR time spectra collected 
		at 0.3\,K and 10\,K in an applied field of 80\,mT, with the 
		respective Fourier transforms shown in (b) and (c). Solid lines 
		are fits to Eq.~\ref{eq:TF_muSR} using two oscillations, which 
		are also shown separately as dash-dotted lines in (b), 
		together with a background contribution. 
		The dashed vertical line indicates the applied magnetic field. 
		In (b), note the clear field-distribution broadening of FLL below $T_c$.}
\end{figure}
The TF-$\mu$SR measurements were carried out in a field of 80\,mT, 
twice the $\mu_{0}H_{c1}(0)$ value. Two representative TF-$\mu$SR spectra 
of Mo$_5$PB$_2$, collected at 0.3\,K and 10\,K (i.e., in the 
superconducting and the normal state) 
are shown in figure~\ref{fig:TF-muSR_T}(a). In the normal state, the 
spectra have essentially no damping, reflecting the uniform 
field distribution, as well as the nomagnetic nature of Mo$_5$PB$_2$. 
Below $T_c$, instead, the significantly enhanced damping occurring in 
the mixed state reflects the inhomogeneous field distribution due to 
the development of FLL~\cite{Yaouanc2011,Amato1997,Blundell1999,Maisuradze2009}. 
This additional SC-related broadening is clearly visible in 
figure~\ref{fig:TF-muSR_T}(b), where the fast-Fourier-transform 
(FFT) spectrum of the corresponding TF-$\mu$SR data 
is shown.
To describe the asymmetric field distribution taking place below $T_c$, 
the $\mu$SR spectra can be modeled by means of the expression: 
\begin{equation}
\label{eq:TF_muSR}
A_\mathrm{TF}(t) = \sum\limits_{i=1}^n A_i \cos(\gamma_{\mu} B_i t + \phi) e^{- \sigma_i^2 t^2/2} +
A_\mathrm{bg} \cos(\gamma_{\mu} B_\mathrm{bg} t + \phi).
\end{equation}
Here $A_i$ 
and $A_\mathrm{bg}$ 
represent the initial muon-spin asymmetries for muons implanted in the 
sample and sample holder, respectively, with the latter giving rise 
to a background signal not undergoing any depolarization. 
$B_i$ and $B_\mathrm{bg}$ are the local fields sensed by the 
implanted muons in the sample and the sample holder (the latter 
normally experiencing the unchanged external field), 
$\gamma_{\mu}/2\pi = 135.53$\,MHz/T is the muon gyromagnetic ratio, 
$\phi$ is a shared initial phase, and $\sigma_i$ is the Gaussian 
relaxation rate of the $i$th component. 

\begin{figure}[!thp]
	\centering
	\includegraphics[width=0.6\textwidth,angle= 0]{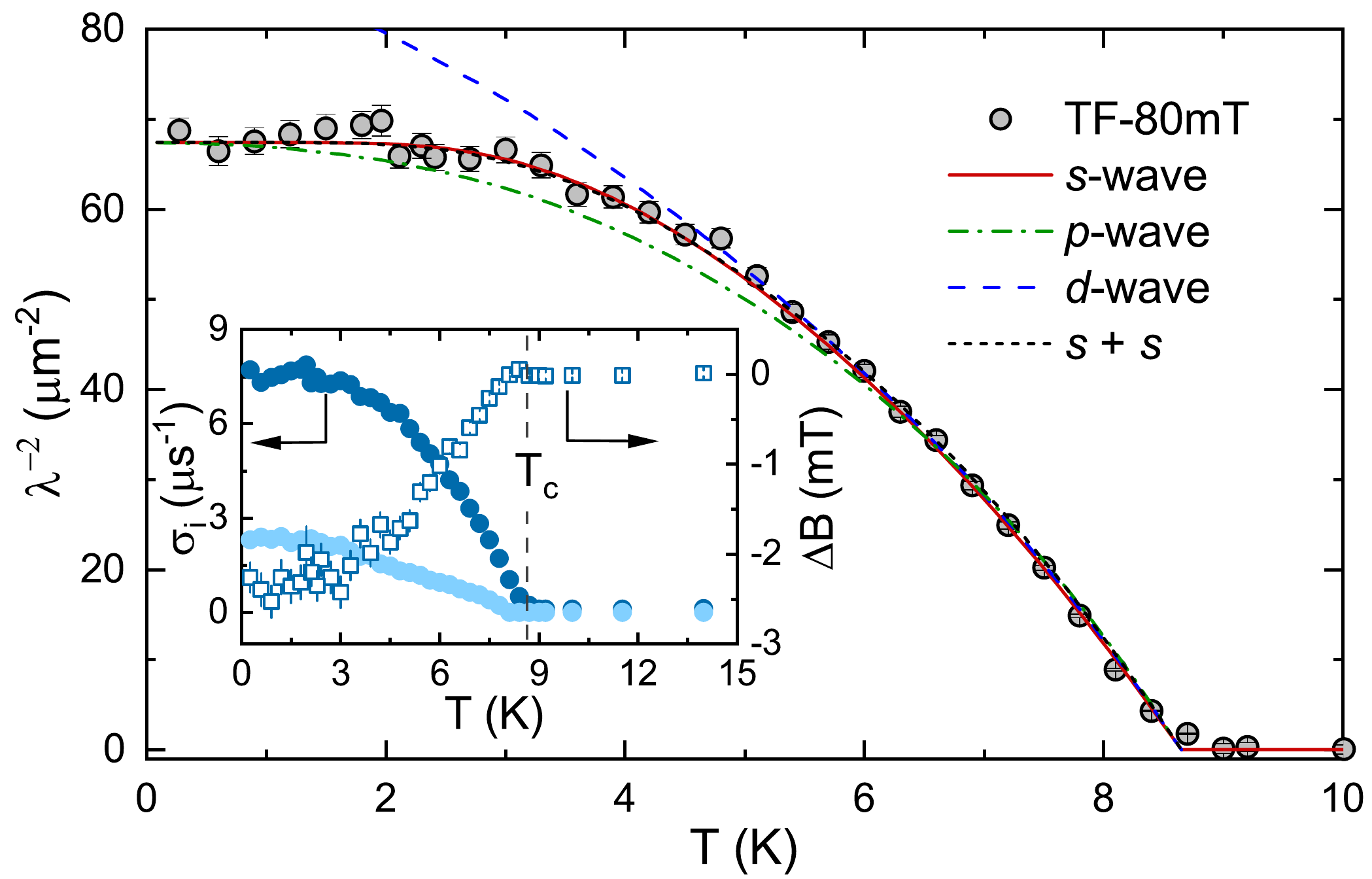}
	\caption{\label{fig:lambda}Superfluid density vs temperature, as 
		determined from TF-$\mu$SR measurements in Mo$_5$PB$_2$ in an 
		applied magnetic field of 80\,mT. The inset shows the temperature 
		dependence of the muon-spin relaxation rate $\sigma_i(T)$ and diamagnetic shift $\Delta B (T) = \langle B \rangle - B_\mathrm{appl.}$. 
		Two $\sigma$s are required to describe the TF-$\mu$SR data [see 
		figure~\ref{fig:TF-muSR_T}(b)].
		The different lines in the main panel represent fits to various models, including \tcr{single-gap $s$-, $p$-, and $d$-wave, and two-gap $s+s$-wave} (see text for details).
	\tcr{Note that, after subtracting a possible Mo$_3$P contribution in quadrature, 
	the resulting data practically overlap with the originally measured $\lambda^{-2}(T)$.} }
\end{figure}

\begin{figure}[ht]
	\centering
	\includegraphics[width=0.55\textwidth,angle=0]{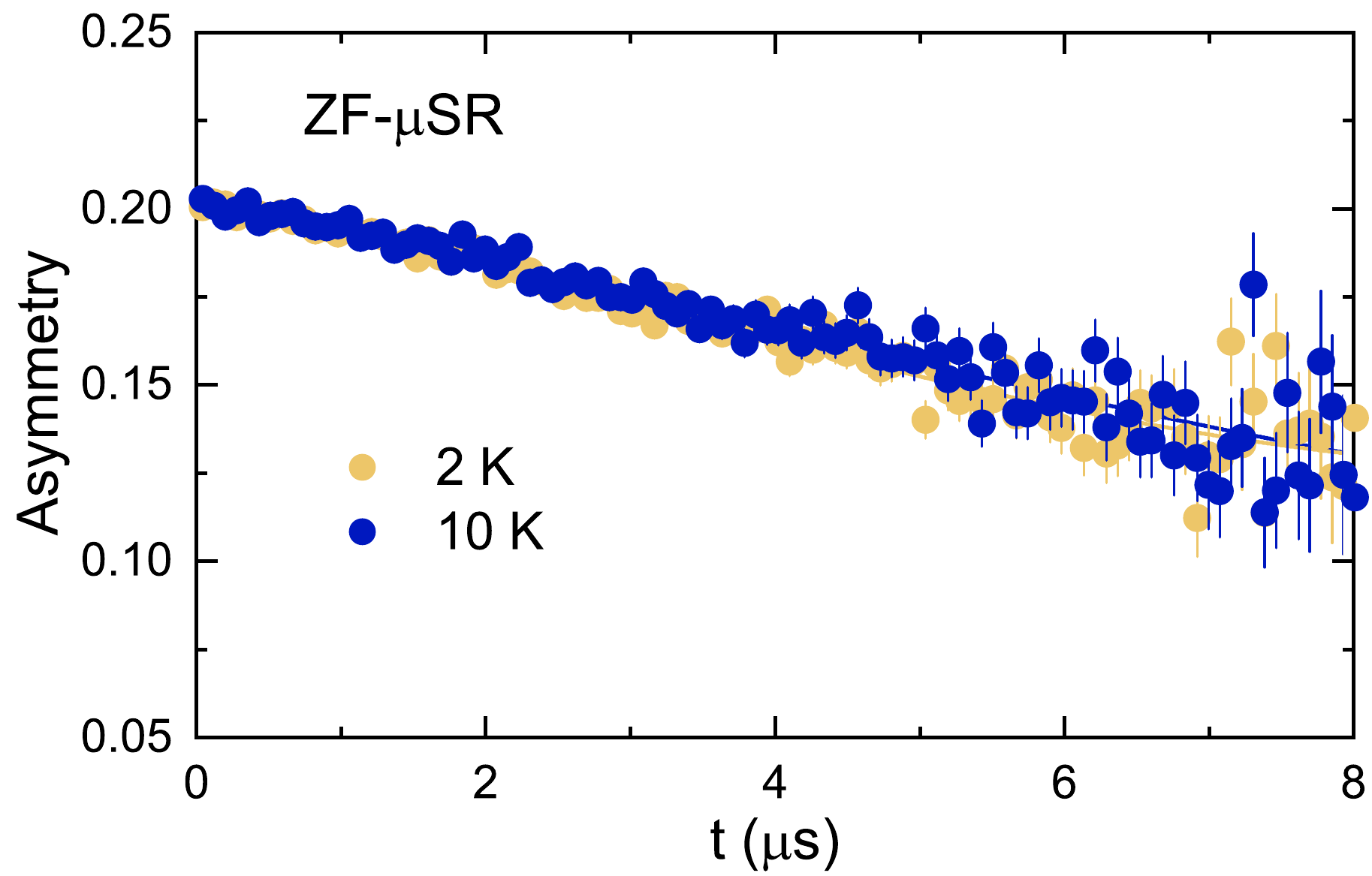}
	\vspace{-2ex}%
	\caption{\label{fig:ZF_muSR}Representative ZF-$\mu$SR spectra in 
		the normal (10\,K) and the superconducting state (2\,K) of Mo$_5$PB$_2$. 
		Solid lines are fits to the equation described in the text. 
		None of the datasets shows noticeable changes with temperature.}
\end{figure}

Generally, the field distribution in the SC state is material dependent: 
the more asymmetric it is, the more components are required to describe it. 
Here we found that, to properly describe the TF-$\mu$SR spectra in the 
superconducting state of Mo$_5$PB$_2$, at least two oscillations 
are required. 
This is illustrated in figure~\ref{fig:TF-muSR_T}(b), where two broad 
peaks, above and below the applied magnetic field (80\,mT), can be 
clearly seen. Both peaks are much broader than the single peak shown 
in figure~\ref{fig:TF-muSR_T}(c), corresponding to the field distribution 
in the normal state. 
The solid lines in figure~\ref{fig:TF-muSR_T} represent fits to Eq.~\ref{eq:TF_muSR} with $n = 2$, while the dash-dotted lines in figure~\ref{fig:TF-muSR_T}(b) evidence 
the single components at 0.3\,K and 
the background signal. The derived Gaussian relaxation rates as a function 
of temperature are summarized in the inset of figure~\ref{fig:lambda}. 
At base temperature (0.3\,K), $\sigma_1 = 7.72(16)$\,$\mu\mathrm{s}^{-1}$ 
and $\sigma_2 = 2.32(10)$\,$\mu\mathrm{s}^{-1}$ reflect the 
$A_1$ and $A_2$ field distributions in figure~\ref{fig:TF-muSR_T}(b), respectively.    
Above $T_c$, the relaxation rate is small and temperature-independent, but below $T_c$ it starts to increase due to the onset of FLL and the increased superfluid density. 
At the same time, also a diamagnetic field shift 
appears below $T_c$, given by $\Delta B (T) = \langle B \rangle - B_\mathrm{appl.}$, 
with $\langle B \rangle = (A_1\,B_1 + A_2\,B_2)/A_\mathrm{tot}$, 
$A_\mathrm{tot} = A_1 + A_2$, and $B_\mathrm{appl.} = 80$\,mT (see inset 
in figure~\ref{fig:lambda}). 
The effective Gaussian relaxation rate can be estimated from 
$\sigma_\mathrm{eff}^2/\gamma_\mu^2 = \sum_{i=1}^2 A_i [\sigma_i^2/\gamma_{\mu}^2 - \left(B_i - \langle B \rangle\right)^2]/A_\mathrm{tot}$~\cite{Maisuradze2009}.
Then, the superconducting Gaussian relaxation rate, encoded in 
the $\sigma_\mathrm{FLL}$ value, can be extracted by subtracting 
the nuclear contribution according to 
$\sigma_\mathrm{FLL} = \sqrt{\sigma_\mathrm{eff}^{2} - \sigma^{2}_\mathrm{n}}$. 
Here, $\sigma_\mathrm{n} \sim 0.11\,\mu\mathrm{s}^{-1}$ is the nuclear relaxation rate, almost constant in our narrow temperature range, 
as confirmed by zero-field (ZF) $\mu$SR data (see figure~\ref{fig:ZF_muSR}).
For small applied magnetic fields ($H_\mathrm{appl}$/$H_{c2}$ $\sim$ 0.04 $\ll$\,1), the magnetic penetration depth $\lambda$ can be calculated from 
$\sigma_\mathrm{sc}^2(T)/\gamma^2_{\mu} = 0.00371\Phi_0^2/\lambda^4(T)$~\cite{Barford1988,Brandt2003}.
%
%
Figure~\ref{fig:lambda} shows the temperature dependent inverse square of the magnetic penetration depth [proportional to the superfluid density, i.e.., $\lambda^{-2}(T) \propto \rho_\mathrm{sc}(T)$] for Mo$_5$PB$_2$.  The  superfluid density $\rho_\mathrm{sc}(T)$ was further analyzed by using 
different models, generally described by:
\begin{equation}
\label{eq:rhos}
\rho_\mathrm{sc}(T) = 1 + 2\, \Bigg{\langle} \int^{\infty}_{\Delta_\mathrm{k}} \frac{E}{\sqrt{E^2-\Delta_\mathrm{k}^2}} \frac{\partial f}{\partial E} \mathrm{d}E \Bigg{\rangle}_\mathrm{FS}. 
\end{equation}
Here, $f = (1+e^{E/k_\mathrm{B}T})^{-1}$ is the Fermi function and $\langle \rangle_\mathrm{FS}$ represents an average over the Fermi surface~\cite{Tinkham1996}. 
$\Delta_\mathrm{k}(T) = \Delta(T) \delta_\mathrm{k}$ is an angle-dependent gap function, where $\Delta$ is the maximum gap value and $\delta_\mathrm{k}$ is the 
angular dependence of the gap, equal to 1, $\cos2\phi$, and $\sin\theta$ 
for an $s$-, $d$-, and $p$-wave model, respectively, with $\phi$ 
and $\theta$ being the azimuthal angles.
The temperature dependence of the gap is assumed to follow $\Delta(T) = \Delta_0 \mathrm{tanh} \{1.82[1.018(T_\mathrm{c}/T-1)]^{0.51} \}$~\cite{Tinkham1996,Carrington2003}, where $\Delta_0$ is the gap value at 0\,K.

\tcr{Four different models, including single-gap $s$-, $p$-, and $d$-wave, and two-gap $s+s$-wave},  
were used to describe the $\lambda^{-2}$$(T)$ dependence. 
For an $s$- or $p$-wave model, the best fits yield the same 
zero-temperature magnetic penetration depth $\lambda_\mathrm{0} = 121(2)$\,nm, 
but different gap values, 1.42(2) and 1.87(2)\,meV, respectively. 
For the $d$-wave model, the estimated $\lambda_\mathrm{0}$ and gap 
value are 104(2)\,nm and 1.75(2)\,meV. 
As can be clearly seen in figure~\ref{fig:lambda}, the significant 
deviation of the $p$- or $d$-wave model from the experimental data 
below 5\,K and the temperature-independent behavior of $\lambda^{-2}(T)$ 
for $T < 1/3T_c \sim 3$\,K strongly suggest a fully-gapped 
superconductivity in Mo$_5$PB$_2$. According to previous studies~\cite{McGuire2016}, two gaps are required to quantitatively describe the specific-heat data (as confirmed also here, see below). 
\tcr{Here, by fixing the weight $w = 0.25$, as determined from the electronic 
specific heat (see below), the two-gap $s+s$-wave model provides almost 
identical results to the single-gap $s$-wave model. The two derived gap 
values $\Delta_0^\mathrm{f} = 1.11(2)$ and $\Delta_0^\mathrm{s} = 1.57(1)$\,meV 
are very similar to those determined from electronic specific heat.}


Since the weight of the second gap is relatively small (0.25--0.3) \tcr{and the gap sizes are not significantly different ($\Delta_0^\mathrm{f}$/$\Delta_0^\mathrm{s}$
	$\sim$ 0.71)}, 
this makes it difficult 
to discriminate between a single- and a two-gap superconductor
based on the temperature-dependent superfluid density alone~\cite{Khasanov2014,Khasanov2020}.  
Nevertheless, as we show further, the two-gap feature of Mo$_5$PB$_2$ 
is clearly reflected also in its field-dependent superconducting 
relaxation rate $\sigma_\mathrm{FLL}(H)$. Since normally the 
different gaps respond differently to an external field, 
$\sigma_\mathrm{FLL}(H)$ exhibits different features in a two-gap superconductor 
compared to a single-gap superconductor.
%
\begin{figure}[!thp]
	\centering
	\includegraphics[width=0.6\textwidth,angle=0]{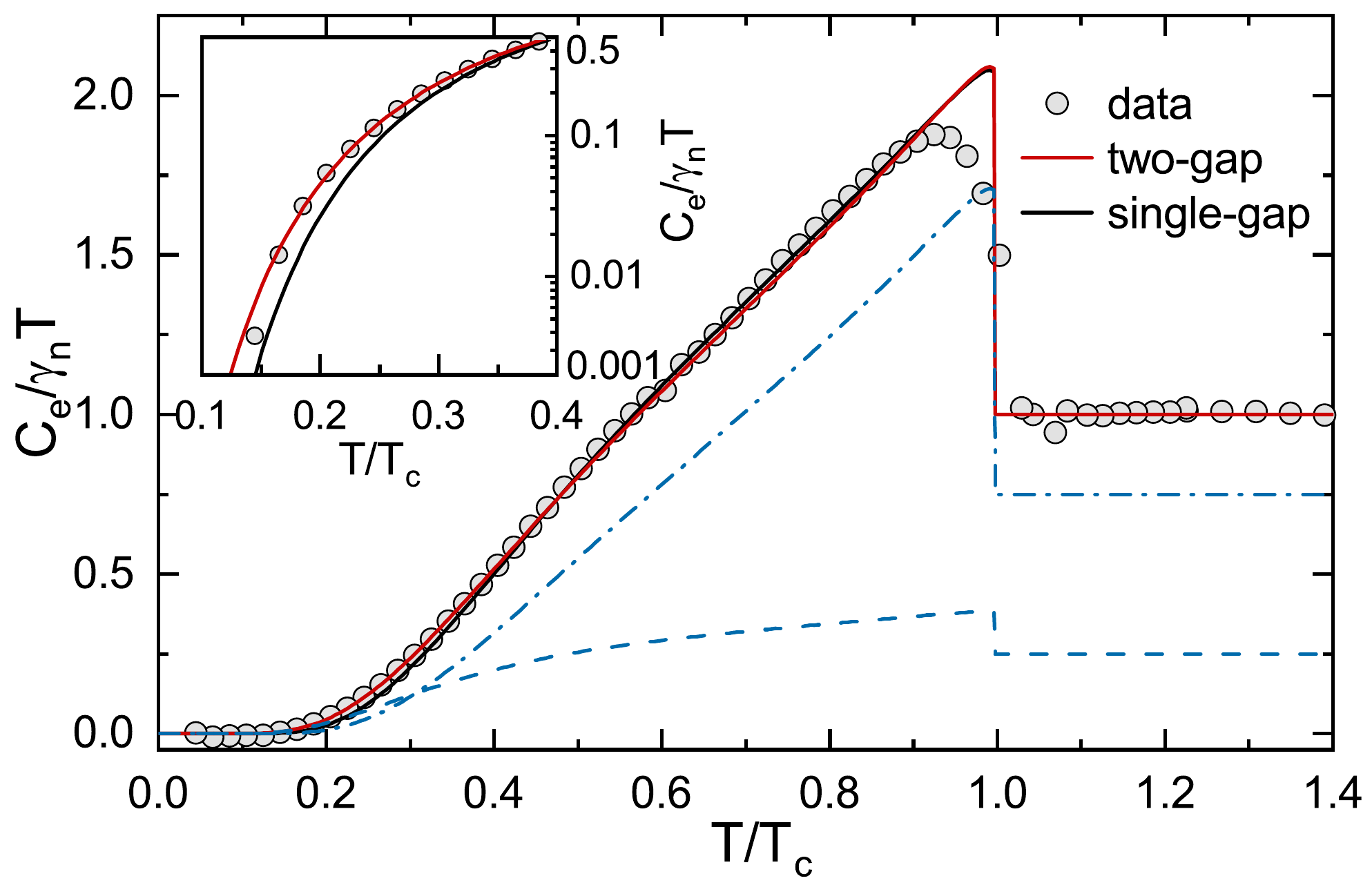}
	\vspace{-2ex}%
	\caption{\label{fig:Cp2}Normalized electronic specific heat 
		$C_\mathrm{e}/\gamma_n T$ of Mo$_5$PB$_2$ as a function of $T$/$T_c$. 
		Inset: enlarged plot of the low-$T$ \tcr{($0.1 \leq T/T_c \leq$ 0.4)} normalized electronic specific heat \tcr{in semi-logarithmic scale}.  
		The solid red and black lines represent the electronic 
		specific heat calculated by considering a fully-gapped $s$-wave model with two gaps or a single gap, respectively.
		The dash-dotted- and dashed blue lines in the main panel represent the individual contributions from the large and small 
		superconducting gaps.
		The goodness of fit is $\chi_r^2 = 1.9$ (two-band model) and  7.9 (single-band model).} 
\end{figure}

To reveal the multigap superconductivity of Mo$_5$PB$_2$, we also 
analyzed the zero-field electronic specific-heat data. After subtracting 
from the raw specific-heat data the phonon contribution \tcr{(see 
details in the inset of figure~\ref{fig:Cp1})}
and the spurious Mo$_3$P contribution (see details in Ref.~\cite{McGuire2016}), 
the resulting electronic specific heat divided by 
the normal-state electronic specific-heat coefficient, i.e., 
$C_\mathrm{e}/\gamma_\mathrm{n}T$, is 
reported in figure~\ref{fig:Cp2}. Since the previous analysis 
of $\lambda^{-2}(T)$ already excluded the 
occurrence of nodes in the 
SC gap, 
the temperature-dependent electronic specific heat was 
analyzed by using a fully-gapped model. 
The solid black line in figure~\ref{fig:Cp2} represents a fit to the  
$s$-wave model with a single gap $\Delta_0 = 1.38(2)$\,meV (i.e., 
equivalent to the standard BCS value 1.76\,k$_\mathrm{B}T_c$).  
It reproduces very well the experimental data above $T/T_c \sim 0.4$.   
Yet, at lower temperatures, the single-gap model shows a less satisfactory 
agreement (see inset).
At the same time, the two-gap model exhibits a much better agreement 
across the full temperature range, in particular for $T/T_c < 0.4$ 
(see inset), reflected in a much smaller $\chi_r^2$ value. 
The solid red line in figure~\ref{fig:Cp2} is a fit to the two-gap $s$-wave model, $C_e(T)/T = wC_e^{\Delta^\mathrm{f}}(T)/T + (1-w)C_e^{\Delta^\mathrm{s}}(T)/T$~\cite{Bouquet2001}. 
Here $C_e^{\Delta^\mathrm{f}}(T)/T$ and $C_e^{\Delta^\mathrm{s}}(T)/T$ are the single-gap 
specific-heat contributions, with $\Delta^\mathrm{f}$ the first- 
(small) and $\Delta^\mathrm{s}$ the second (large) gap, 
and $w$ the relative weight. 
The two-gap model gives $\Delta_0^\mathrm{f} = 1.02(2)$\,meV, 
$\Delta_0^\mathrm{s} = 1.49(2)$\,meV, and $w = 0.25$, the two superconducting 
gap values being consistent with previous results~\cite{McGuire2016}. 
The large-gap value, as well as the gap value determined from 
TF-$\mu$SR, are both greater than $\Delta_\mathrm{BCS}$
expected from the BCS theory in the weak-coupling regime, hence 
indicating \emph{strong-coupling} superconductivity in Mo$_5$PB$_2$.

\subsection{\label{ssec:twoband}Field-dependent measurements: Evidence of multigap superconductivity}
To get further insight into the multigap SC revealed by zero-field electronic specific heat, we also carried out a series of measurements (including TF-$\mu$SR, heat capacity, magnetization, and electrical resistivity) at different magnetic fields. The later three were also used to determine the upper critical field $H_{c2}(T)$. 

%
\begin{figure}[!bhp]
	\centering
	\includegraphics[width=0.61\textwidth,angle=0]{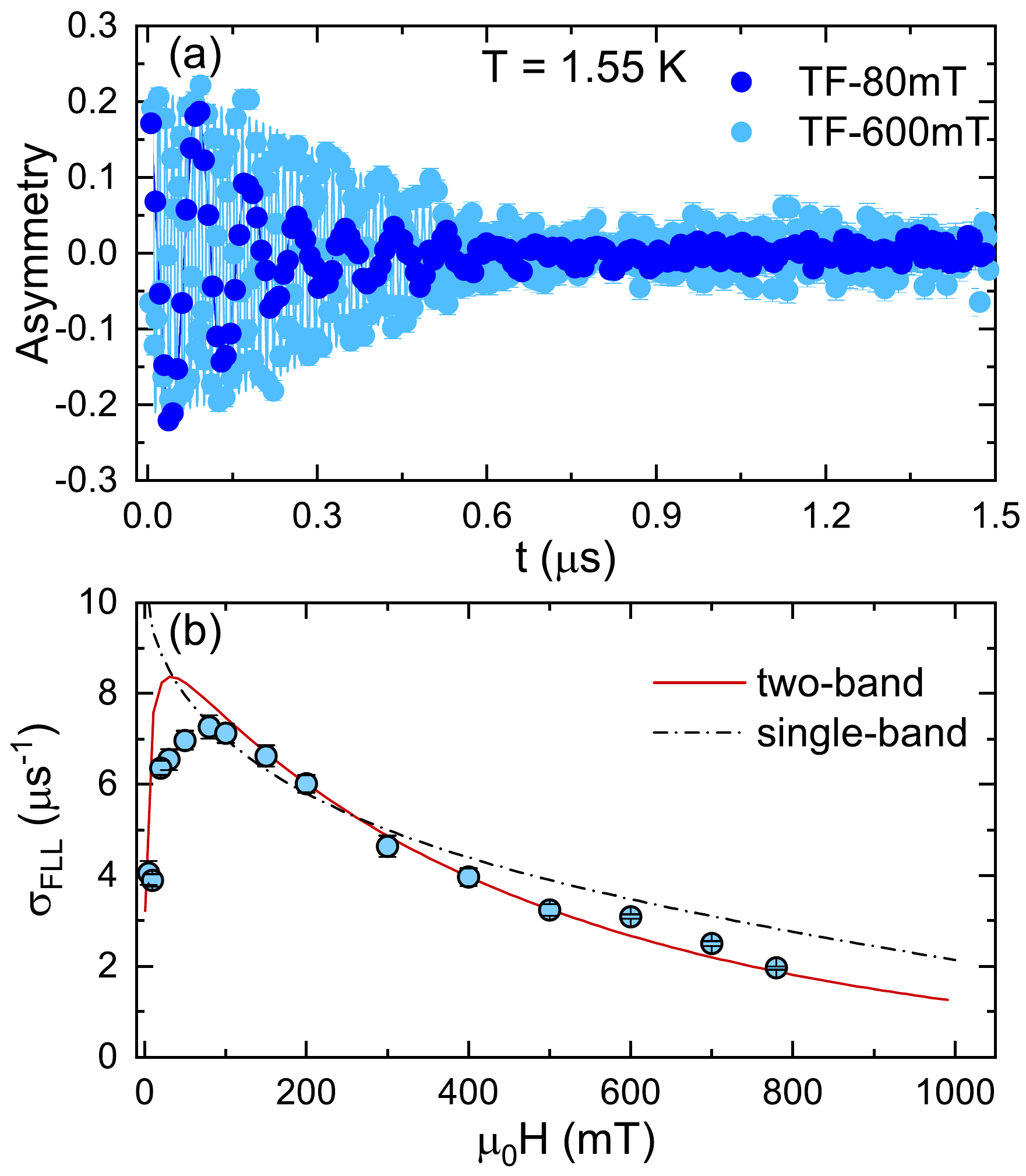}
	\caption{\label{fig:lambda2}(a) TF-$\mu$SR time spectra for Mo$_5$PB$_2$ measured in the superconducting state ($T$ = 1.55\,K) in a field of 80 and 600\,mT. (b) Field-dependent superconducting Gaussian relaxation rate $\sigma_\mathrm{FLL}(H)$. The solid and dash-dotted lines	represent 
		fits to two-band and single-band models, respectively. The poor 
		agreement between theory and experiment below $H_{c1}$ is 
		due to the Meissner effect.} 
\end{figure}
%

\begin{figure}[!htp]
	\centering
	\includegraphics[width = 0.6\textwidth]{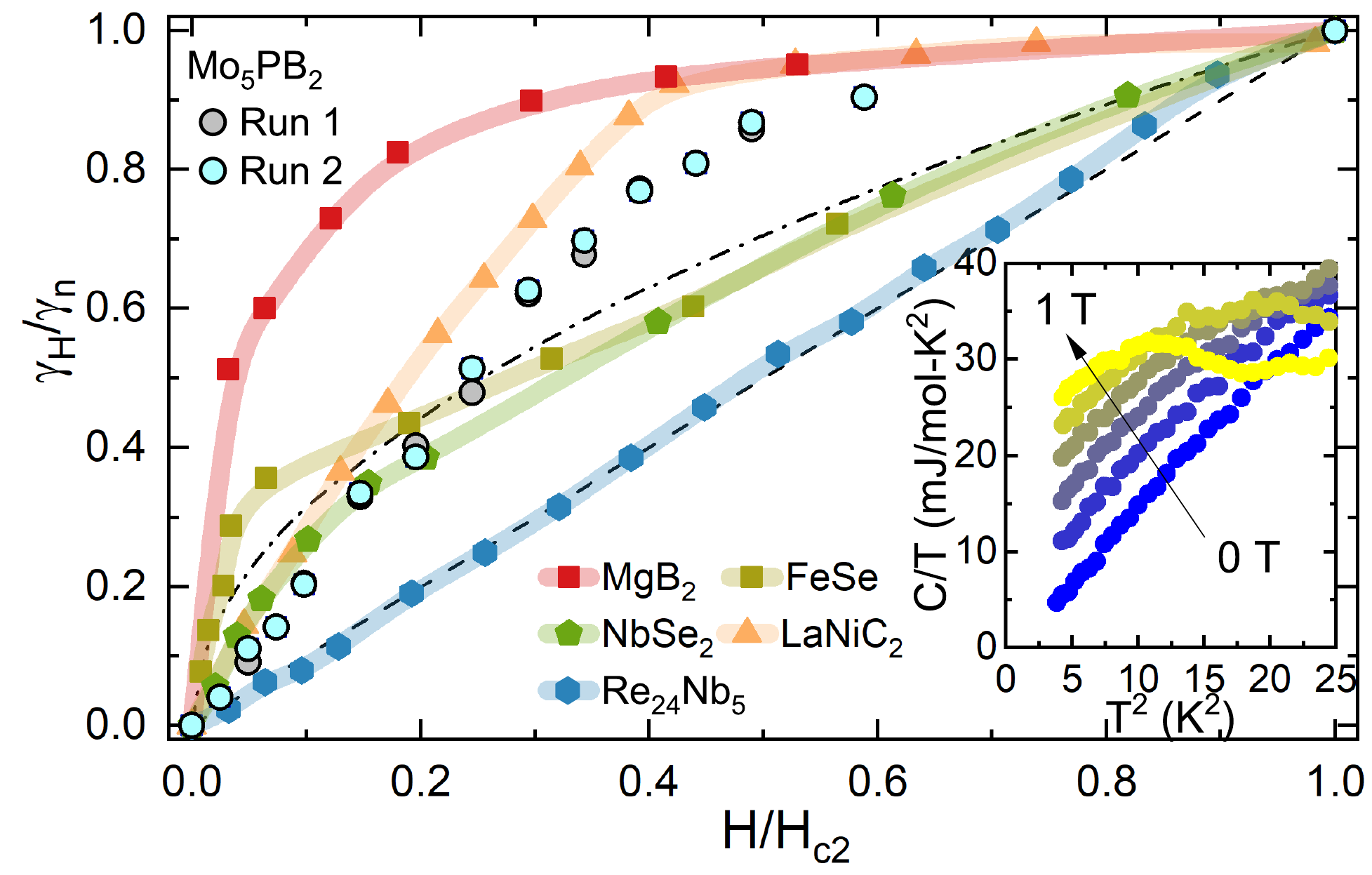}
	\caption{\label{fig:Cp3}Normalized specific-heat coefficient 
		$\gamma_\mathrm{H}$/$\gamma_\mathrm{n}$ vs reduced magnetic 
		field $H/H_{c2}(0)$ for Mo$_5$PB$_2$. 
		At a given applied field, $\gamma_\mathrm{H}$ is obtained as the 
		linear extrapolation of $C/T$ vs $T^2$ in the superconducting 
		state to zero temperature (see inset).
		The dashed and dash-dotted lines represent the $\gamma(H)$ expected 
		for a single-gap model with isotropic or line nodal gap structure, respectively. 
		The data for the reference samples are	adopted from Refs.~\cite{Bouquet2001a,Chen2013,Chen2017,TianReNb2018,Huang2007}.}
\end{figure}

\emph{$\sigma_\mathrm{FLL}$ vs\ $H$}. 
TF-$\mu$SR measurements at different magnetic fields (up to 780\,mT) 
were performed in the superconducting state
of Mo$_5$PB$_2$. As an example, the TF-$\mu$SR spectra collected at 80 and 600\,mT are shown in figure~\ref{fig:lambda2}(a). 
Again the spectra were analyzed using the model described 
by Eq.~\ref{eq:TF_muSR}. The resulting superconducting Gaussian 
relaxation rates $\sigma_\mathrm{FLL}$ versus the applied magnetic field are summarized in 
figure~\ref{fig:lambda2}(b). In case of a single-gap superconductor, 
$\sigma_\mathrm{FLL}(H)$ generally follows $\sigma_\mathrm{FLL} = 0.172 \frac{\gamma_{\mu} \Phi_0}{2\pi}(1-h)[1+1.21(1-\sqrt{h})^3]\lambda^{-2}$~\cite{Barford1988,Brandt2003}, where $h = H_\mathrm{appl}$/$H_{c2}$, 
with $H_\mathrm{appl}$ being the applied magnetic field. 
By fixing $\mu_0H_{c2}$ = 1.77\,T (at 1.55\,K) (see figure~\ref{fig:Hc2}), 
the single-band model clearly deviates from the experimental data at 
magnetic fields above 300\,mT [see dash-dotted line in 
figure~\ref{fig:lambda2}(b)].   
In a two-band model, each band is 
characterized by its own coherence length 
[i.e., $\xi^\mathrm{f}$ (first) and $\xi^\mathrm{s}$ (second)] 
and a weight $w$ [or ($1-w$)], accounting for the relative contribution 
of each band 
to the total $\sigma_\mathrm{FLL}$ and, hence, to the superfluid 
density~\cite{Serventi2004,Khasanov2014}. 
By fixing $w = 0.25$, as estimated frome 
electronic specific-heat data (figure~\ref{fig:Cp2}), the 
two-band model [solid red line in figure~\ref{fig:lambda2}(b)] 
is in better agreement with the experiment
and provides $\lambda_0$ = 99(2)\,nm, $\xi^\mathrm{f}$ = 18.5(5)\,nm, and 
$\xi^\mathrm{s}$ = 13.2(2)\,nm. 
The upper critical field of 1.89(5)\,T, calculated from the coherence 
length of the second band, $\mu_0H_{c2} = \Phi_0/(2\pi\xi^{2})$, 
is also comparable to the upper critical field determined from 
bulk measurements. The \emph{virtual} upper critical field  
$\mu_0H_{c2}^\ast$ = 0.96(5)\,T, calculated from the coherence 
length of the first band $\xi^\mathrm{f}$, is in good agreement with the field 
value where both $H_{c2}(T)$ (figure~\ref{fig:Hc2}) and $\gamma_\mathrm{H}(H)$ (figure~\ref{fig:Cp3}) show a flex or change the slope, respectively.
\begin{figure}[!htp]
	\centering
	\includegraphics[width=0.5\textwidth,angle= 0]{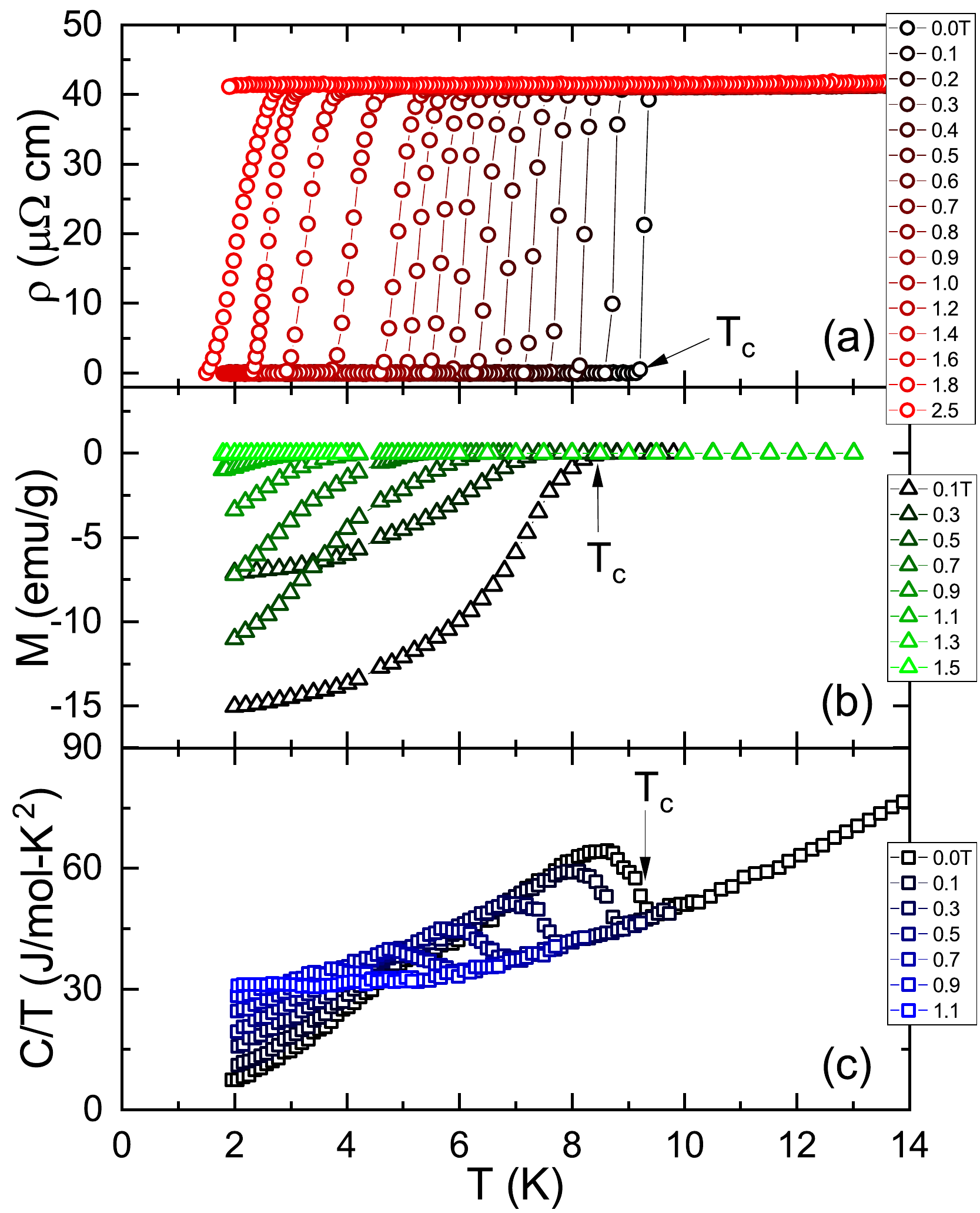}
	\caption{\label{fig:Hc2_determ}Temperature-dependent (a) electrical 
	resistivity $\rho(T,H)$, (b) magnetization $M(T,H)$, and (c) 
	specific-heat data $C(T,H)/T$, collected at various magnetic fields 
	up to 2.5\,T. For the $\rho(T,H)$ measurements, $T_c$ was defined 
	as the onset of zero resistivity; while for the $M(T,H)$ and $C(T,H)/T$ 
	measurements, $T_c$ was defined as the onset and the midpoint of 
	the superconducting transition, respectively. \tcr{All $T_c(0)$ 
	values are marked by arrows}.}
\end{figure}
%

%
\begin{figure}[!htp]
	\centering
	\includegraphics[width=0.6\textwidth,angle= 0]{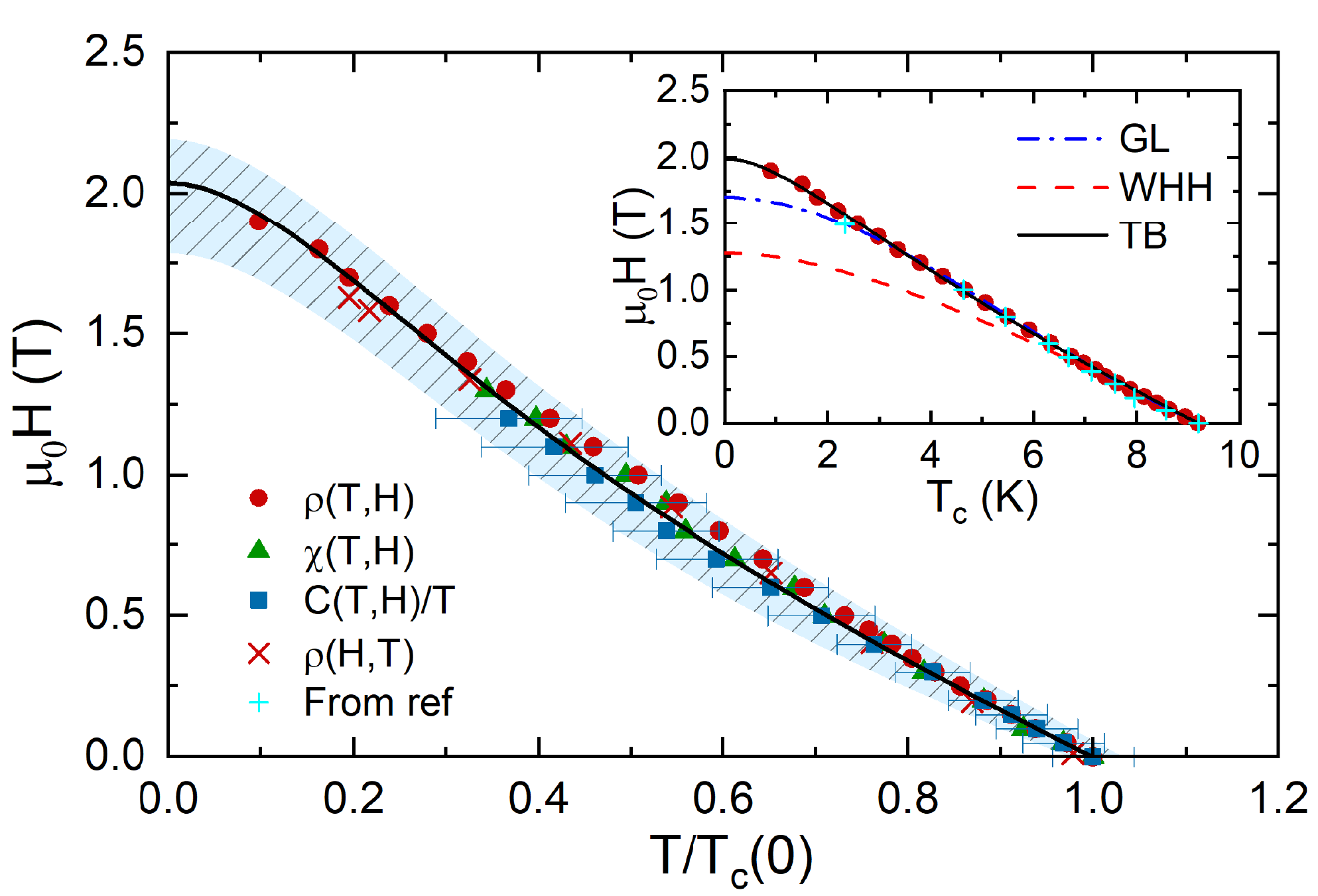}
	\caption{\label{fig:Hc2}(a) Upper critical field $\mu_{0}H_{c2}$ vs  
		reduced transition temperature $T_c/T_c(0)$ for Mo$_5$PB$_2$. 
		The $T_c$ values were determined from measurements shown in 
		figure~\ref{fig:Hc2_determ}. 
		Inset shows the critical field vs $T_c$, as determined 
		from $\rho(T,H)$ and 
		data taken from Ref.~\cite{McGuire2016}. 
		Three different fits, using the GL- (dash-dotted line), WHH- 
		(dashed line), and TB model (solid line) are also shown 
		in the inset. 
		The error bars are determined as the superconducting transition 
		widths $\Delta$$T_c$ in the specific-heat data. 
		The shaded region indicates the upper- and lower $H_{c2}$ limits, as determined using the two-band model.}
\end{figure}
%

\emph{Upper critical field}. The upper critical field 
$H_{c2}$ of Mo$_5$PB$_2$ was determined from measurements of 
the electrical resistivity $\rho(T,H)$, magnetization $M(T,H)$, and specific heat $C(T,H)/T$ under various applied magnetic fields up to 2.5\,T, as 
shown in figure~\ref{fig:Hc2_determ}(a) to (c). 
Under applied field, the superconducting transition shifts towards 
lower temperatures and becomes broader.   
The $H_{c2}$ values, determined using different techniques, 
are highly consistent and are summarized in figure~\ref{fig:Hc2} as a 
function of the reduced temperature $T_c$/$T_c$(0) [here, $T_c$(0) is 
the transition temperature in zero field]. 
The $H_{c2}(T)$ was analyzed by means of Ginzburg-Landau (GL)~\cite{Zhu2008}, Werthamer-Helfand-Hohenberg (WHH)~\cite{Werthamer1966}, and two-band (TB) models~\cite{Gurevich2011}. 
As shown in the inset of figure~\ref{fig:Hc2}, the GL model reproduces 
the experimental data up to $\mu_0H \sim 1.4$\,T, 
while the WHH model stops already at 0.5\,T. At higher magnetic fields, both 
models show large deviations, leading to underestimated values of 
$\mu_0 H_{c2}^\mathrm{GL}(0) =$ 1.7(1)\,T and 
$\mu_0 H_{c2}^\mathrm{WHH}(0)=$ 1.3(1)\,T. 
Such discrepancy most likely hints at multiple superconducting gaps 
in Mo$_5$PB$_2$, as evidenced also by the positive curvature of 
$H_{c2}(T)$ at low fields, a typical feature of multigap 
superconductors, as e.g., MgB$_2$~\cite{Muller2001,Gurevich2004} or  
Lu$_2$Fe$_3$Si$_5$~\cite{Nakajima2012}. 
As shown in figure~\ref{fig:Hc2}, around $T_c/T_c$(0) $\sim$ 0.5 
($\mu_0H$ $\sim$ 0.93\,T), $H_{c2}(T)$ undergoes a clear change 
in curvature, which coincides with $\mu_0H_{c2}^\ast = 0.96$\,T 
of the first superconducting band 
(see figure~\ref{fig:lambda2}). The remarkable agreement of the TB model 
with the experimental data across the full temperature range is clearly seen in 
figure~\ref{fig:Hc2}, from which we find $\mu_0H_{c2}^\mathrm{TB}(0) =$ 2.0(2)\,T and $\xi(0)$ = 12.8(6)\,nm. 
Note that the $T_c$ and $\mu_0$$H_{c2}$ values of the spurious 
Mo$_3$P phase~\cite{Shang2019} are both much smaller than 
those of Mo$_5$PB$_2$. Consequently, the 
two-gap feature of $H_{c2}(T)$ is intrinsic to Mo$_5$PB$_2$.
The lower critical field $\mu_{0}H_{c1}$ is related to the magnetic penetration 
depth $\lambda$ and the coherence length $\xi$ via $\mu_{0}H_{c1} = (\Phi_0 /4 \pi \lambda^2)[$ln$(\kappa)+ 0.5]$, 
where $\kappa$ = $\lambda$/$\xi$ is the GL parameter~\cite{Brandt2003}.
By using $\mu_{0}H_{c1} = 30.4(4)$\,mT and $\mu_{0}H_{c2} = 2.0(2)$\,T, 
the resulting magnetic penetration depth $\lambda_\mathrm{GL}$ = 122(2)\,nm, 
is almost identical to the experimental value 121(2)\,nm determined 
from TF-$\mu$SR data (see Sec.~\ref{ssec:TF_muSR}). 
A large GL parameter, $\kappa \sim 9.5$, clearly indicates that 
Mo$_5$PB$_2$ is a type-II superconductor.  
%
%

\emph{$\gamma_\mathrm{H}$ vs $H$}. The multigap SC of Mo$_5$PB$_2$ 
is further confirmed by the field-dependent electronic specific 
heat coefficient $\gamma_\mathrm{H}(H)$. 
Since the virtual $\mu_0H_{c2}^\ast$ corresponds to the critical 
field which suppresses the small superconducting gap, we expect also 
$\gamma_\mathrm{H}(H)$ to change its slope around $\mu_0H_{c2}^\ast$. 
The normalized $\gamma_\mathrm{H}/\gamma_\mathrm{n}$ values vs 
the reduced magnetic field $H/H_{c2}(0)$ are shown in 
figure~\ref{fig:Cp3} (here $\gamma_\mathrm{n}$ is the zero-field 
normal-state value). 
Note that, the field dependence of $\gamma_\mathrm{H}/\gamma_\mathrm{n}$ 
measured at 0.4\,K exhibits similar features to that evaluated at zero 
temperature.
For Mo$_5$PB$_2$, $\gamma_\mathrm{H}(H)$ clearly deviates 
from the linear field dependence expected for fully-gapped 
superconductors with a single gap, as e.g., Re$_{24}$Nb$_{5}$
(dashed line)~\cite{Caroli1964,TianReNb2018}, or from the 
square-root dependence $\sqrt{H}$ (dash-dotted line), expected for 
nodal superconductors~\cite{Volovik1993,Wen2004}. 
Instead, Mo$_5$PB$_2$ exhibits similar features to other well known
multigap superconductors, as e.g., FeSe~\cite{Chen2017}, 
MgB$_2$~\cite{Bouquet2001a}, or NbSe$_2$~\cite{Huang2007}. 
The $\gamma_\mathrm{H}(H)$ curve of Mo$_5$PB$_2$ 
(scatter plot)
exhibits a significant change of slope around $H/H_{c2}(0) 
\sim$ 0.45 (i.e., $\mu_0H \sim$ 0.9\,T), which is highly consistent with $\mu_0H_{c2}^\ast$. 

%
%
%
\subsection{\label{ssec:ZF_muSR} Zero-field $\mu$SR}
We also performed ZF-$\mu$SR measurements in both the normal- and 
the superconducting states of Mo$_5$PB$_2$. As shown in figure~\ref{fig:ZF_muSR}, 
neither coherent oscillations nor fast decays could be identified in 
the spectra collected above (12\,K) and below $T_c$ (2\,K), hence 
implying the lack of any magnetic order or fluctuations. 
The weak muon-spin relaxation in absence of an external magnetic 
field is mainly due to the randomly oriented nuclear moments, 
which can be modeled by a Gaussian Kubo-Toyabe relaxation function, 
$G_\mathrm{KT} = [\frac{1}{3} + \frac{2}{3}(1 -\sigma_\mathrm{ZF}^{2}t^{2})\,\mathrm{e}^{-\sigma_\mathrm{ZF}^{2}t^{2}/2}]$
~\cite{Kubo1967,Yaouanc2011}.
Here, $\sigma_\mathrm{ZF}$ is the zero-field Gaussian relaxation rate. 
The solid lines in figure~\ref{fig:ZF_muSR} represent fits to 
the data by considering also an additional zero-field Lorentzian 
relaxation $\Lambda_\mathrm{ZF}$, i.e., $A_\mathrm{ZF}(t) = A_\mathrm{s} G_\mathrm{KT} \mathrm{e}^{-\Lambda_\mathrm{ZF} t} + A_\mathrm{bg}$. 
The relaxations in the normal- and the superconducting states are almost 
identical, as confirmed
by the practically overlapping ZF-$\mu$SR 
spectra above and below $T_c$. This lack of evidence for an additional 
$\mu$SR relaxation below $T_c$ excludes a possible time-reversal 
symmetry breaking in the superconducting state of Mo$_5$PB$_2$.

\subsection{\label{ssec:Dis}Electronic band-structure calculations and discussion}
Apart from the zero-field electronic specific heat (see figure~\ref{fig:Cp2} 
and Ref.~\cite{McGuire2016}), at a microscopic level, 
the multigap superconductivity of Mo$_5$PB$_2$ was also 
probed by field-dependent $\mu$SR relaxation $\sigma_\mathrm{sc}(H)$ 
in the superconducting state (figure~\ref{fig:lambda2}). 
Macroscopically, further evidence was brought by the temperature-dependent 
upper critical field $\mu_0H_{c2}(T)$ (figure~\ref{fig:Hc2}) and the 
field-dependent electronic specific heat coefficient $\gamma_\mathrm{H}(H)$ 
(figure~\ref{fig:Cp3}). 
Our data clearly indicate that Mo$_5$PB$_2$ is a multiband 
superconductor with two distinct superconducting gaps, both opening 
below $T_c$.
\tcr{Although extraneous phases, such as Mo$_3$P, might potentially 
	influence the reported results, we found that their influence is negligible (both qualitatively and quantitatively).}
Below we present that the multigap SC is also 
supported by electronic band-structure calculations.

\begin{figure}[ht]
	\centering
	\includegraphics[width = 0.47\textwidth]{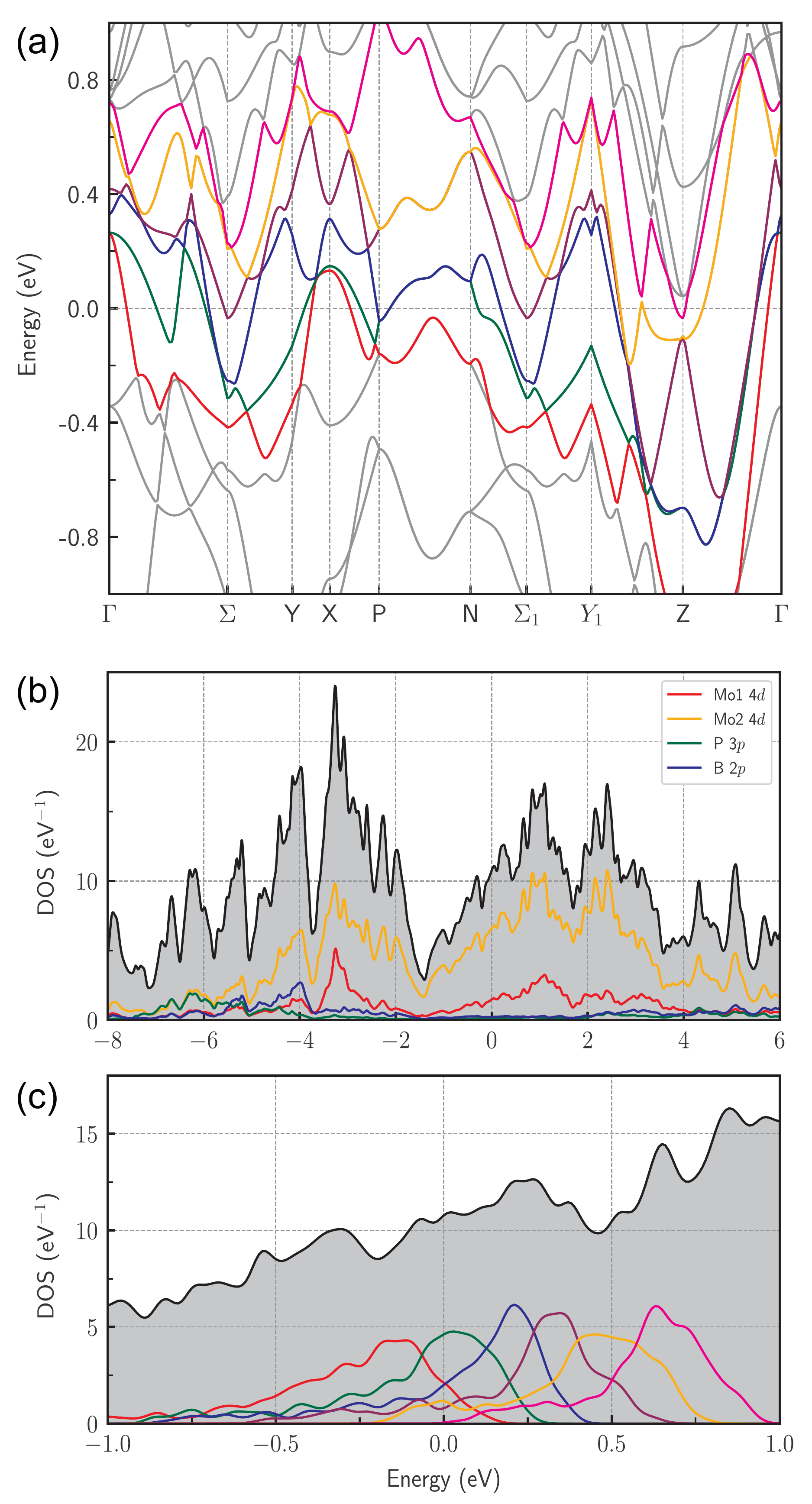}
	\caption{(a) Electronic band structure of Mo$_5$PB$_2$, 
		calculated by ignoring the spin-orbit coupling. 
		The various bands which cross the Fermi level are plotted 
		in different colors. Total- and partial density of states 
		near the Fermi level for (b) different atoms and (c) the 
		six different bands. } 
	\label{fig:DOS}
\end{figure}

As can be seen in figure~\ref{fig:DOS}, six different bands are 
identified to cross the Fermi level. Among these, bands 1 (red-), 
2 (green-) and 3 (blue line), all stemming primarily from the 
Mo 4$d$ orbitals, contribute significantly to the 
density of states at the Fermi level (see table~\ref{tab:VF}). 
We expect the multiband features of Mo$_5$PB$_2$ to be 
closely related to the different site symmetries of Mo atoms in 
the unit cell, namely, Mo1 (4$c$) and Mo2 (16$l$). 
According to band-structure calculations, the contribution of 
16$l$-Mo atoms to the DOS is preponderant compared to that of 
4$c$-Mo atoms [see figure~\ref{fig:DOS}(b)].
%
%
%
\begin{table}[!bht]
	\caption{\label{tab:VF}Calculated Fermi velocity $v_\mathrm{F}$ for 
	the different bands near the Fermi level and the band contributions 
	to the total DOS. Here $v_\mathrm{F}$ is in 10$^5$\,m/s units.} 
	\begin{indented}
		\lineup
		\item[]\begin{tabular}{@{}*{7}{c}}
			\br
			\multicolumn{1}{l}{Index} &\multicolumn{1}{c}{DOS(\%)}&%
			\multicolumn{4}{c}{$v_\mathrm{F}$($\Gamma$-$X$)} &%
			\multicolumn{1}{c}{$v_\mathrm{F}$($\Gamma$-$Z$)} \\ 
			\mr 
			1  & 19.72  &  5.82   & 6.39   &  --     & --      & 7.92 \\
			2  & 42.84  &  2.59   & 8.03   & 3.48    & 2.61    & 5.32 \\ 
			3  & 18.58  &  --     & --     & 5.29    & 4.43    & 5.32 \\ 
			4  & \07.35 &  --     & --     & 2.89    & 2.10    & 7.05 \\ 
			5  & 10.86  &  --     & --     &  --     & --      & 2.88 \\ 
			6  & \00.48 &  --     & --     &  --     & --      & 5.49 \\
			\br
		\end{tabular}
	\end{indented}
\end{table}
%
%
%
The Fermi velocities $v_\mathrm{F}$ of these bands, 
calculated along the $\Gamma$-$X$ and $\Gamma$-$Z$ directions, are 
summarized in table~\ref{tab:VF}. 
Considering also the relative weights, the average $v_\mathrm{F}$ is 
comparable to the experimental value
(see table~\ref{tab:parameter}).

The deviation of $\sigma_\mathrm{sc}(H)$ (figure~\ref{fig:lambda2}) 
from a single-band model and the appearance of an upward curvature 
in the upper critical-field data (figure~\ref{fig:Hc2}), both reflect 
the occurrence of two distinct coherence lengths for two different bands, 
here leading to distinct upper critical fields. The Ginzburg-Landau 
coherence length determined from the upper critical field, 
$\xi(0) = \sqrt{\Phi_0/(2\pi H_{c2})}$, is proportional to the BCS 
coherence length $\xi_0$, i.e., $\xi(0) = 0.855\sqrt{\xi_0 l_e}$ ~\cite{Tinkham1996}. At zero temperature, the 
BCS coherence length is also related to the superconducting energy 
gap $\Delta_0$ and the Fermi velocity $v_\mathrm{F}$, i.e.,  $\xi_0 = \hbar v_\mathrm{F}/\pi \Delta_0$. Therefore, for a multigap superconductor such 
as Mo$_5$PB$_2$, 
$v_\mathrm{F}^\mathrm{f}/v_\mathrm{F}^\mathrm{s} = \xi_0^\mathrm{f}\Delta_0^\mathrm{f}/\xi_0^\mathrm{s}\Delta_0^\mathrm{s}$.
According to the zero-field electronic specific-heat results, 
$\Delta_0^\mathrm{f}/\Delta_0^\mathrm{s} = 1.02/1.49$, while the analysis 
of $\sigma_\mathrm{sc}(H)$ with a two-band model yields 
$\xi^\mathrm{f}$(1.5\,K) = 18.5(5)\,nm and $\xi^\mathrm{s}$(1.5\,K) = 13.2(2)\,nm. 
Assuming $\xi^\mathrm{f}/\xi^\mathrm{s} = \xi_0^\mathrm{f}/\xi_0^\mathrm{s}$, 
we find $v_\mathrm{F}^\mathrm{f}/v_\mathrm{F}^\mathrm{s}$ = 0.95, which is highly 
consistent with the theoretical estimates reported in table~\ref{tab:VF}. 
For instance, along the $\Gamma$-$Z$ direction, the dominant 
bands (1, 2, and 3) show very similar $v_\mathrm{F}$ values.


\begin{table}[!bht]
	\caption{\label{tab:parameter}Normal- and superconducting-state properties of Mo$_5$PB$_2$. 
		The London penetration depth $\lambda_\mathrm{L}$, 
		effective mass $m^{\star}$,
		carrier density $n_\mathrm{s}$, BCS coherence length $\xi_0$, electronic 
		mean-free path $l_e$, Fermi velocity $v_\mathrm{F}$, and effective Fermi temperature $T_\mathrm{F}$ are also listed.}
        \begin{indented}
		\item[]\begin{tabular}{lcp{8mm}lc}
			\br
			Property                               & Value (uncert.)             && Property                            & Value (uncert.) \\ 
			\mr	
			$T_c$$^{\rm a}$                        & 9.20(2)\,K                  &&  $\mu_0H_{c2}$                         & 2.0(2) \,T  \rule{0pt}{2.6ex} \\  
			$\rho_0$                               & 41.1(2)\,$\mu\Omega$cm      &&  $\mu_0H_{c2}^\ast$                    & 0.96(5)\,T \\ 
			$\Theta_\mathrm{D}^\mathrm{R}$         & 236(5)\,K                   &&  $\xi(0)$                              & 12.8(6)\,nm     \\ 
			$\mu_0H_{c1}$                          & 30.4(4)\,mT                 &&  $\kappa$                              & 9.5(5)       \\
			$\mu_0H_{c1}^{\mu\mathrm{SR}}$         & 30.8(6)\,mT                 &&  $\lambda_0$                           & 121(2)\,nm \\
			$\gamma_n$                             & 22.3(2)\,mJ/mol-K$^2$       &&  $\lambda_0$$^{\rm b}$                 & 99(2)\,nm      \\
			$\Theta_\mathrm{D}^\mathrm{C}$         & 300(5)\,K                   &&  $\lambda_\mathrm{GL}$                 & 122(2)\,nm  \\	
			$\Theta_\mathrm{E}^\mathrm{C}$         & 530(5)\,K                   &&  $\lambda_\mathrm{L}$                  & 54(4)\,nm    \\
			$\Delta_0$($p$-wave)($\mu$SR)        & 1.87(2)\,meV                  &&  $l_e$                                 & 8.4(6)\,nm  \\
			$\Delta_0$($d$-wave)($\mu$SR)         & 1.76(2)\,meV                 &&  $\xi_0$                               & 34.2(6)\,nm \\
			$\Delta_0$($s$-wave)($\mu$SR)         & 1.42(1)\,meV                 &&	 $m^{\star}$                          & 5.7(5)\,$m_e$   \\
			$w$                                   & 0.25                         &&  $n_\mathrm{s}$           &  5.9(7) $\times$ 10$^{28}$\,m$^{-3}$  \\	
			\tcr{$\Delta_0^\mathrm{f}$($\mu$SR)$^{\rm c}$}         & 1.11(2)\,meV                 &&  $v_\mathrm{F}$           &  2.3(2) $\times$ 10$^5$\,ms$^{-1}$  \\  
	     	\tcr{$\Delta_0^\mathrm{s}$($\mu$SR)$^{\rm c}$}          & 1.57(1)\,meV                 &&  $T_\mathrm{F}$           &  2.1(2) $\times$ 10$^4$\,K \\    
			$\Delta_0^\mathrm{f}(C)$$^{\rm c}$                & 1.02(2)\,meV                 &&                           &                                         \\  
			$\Delta_0^\mathrm{s}(C)$$^{\rm c}$                & 1.49(2)\,meV                 &&                           &              \\    
			\br                     
		\end{tabular}
	    \item[] $^{\rm a}$ Similar values were determined via electrical resistivity, magnetic susceptibility, and heat-capacity measurements.
 	    \item[] $^{\rm b}$ Derived from a two-band-model fit to $\sigma_\mathrm{FLL}(H)$ at 1.5\,K.	
	 	\item[] $^{\rm c}$ \tcr{Derived from a two-gap model analysis}.
        \end{indented}
\end{table}

\vspace{7pt}
\section{\label{ssec:Sum} Conclusion}
In summary, we studied the multigap superconductor Mo$_5$PB$_2$ by means of electrical resistivity, 
magnetization, heat capacity, and $\mu$SR, as well as via numerical calculations.
The temperature dependence of the zero-field electronic specific heat 
and superfluid density reveal a nodeless superconductivity, well described by an isotropic $s$-wave model.
The mul\-ti\-gap features, originally inferred 
from zero-field specific-heat data are further supported by the field-dependent 
electronic specific-heat coefficient and the superconducting Gaussian relaxation rate, as well as by the temperature dependence of the upper critical field. The lack of spontaneous magnetic fields below $T_c$ indicates that time-reversal symmetry is preserved in the superconducting state of Mo$_5$PB$_2$. 
By combining the extensive experimental results presented 
here with numerical band-structure calculations, we can provide 
solid evidence for multigap superconductivity in Mo$_5$PB$_2$.

\ack{The authors thank Chien-Lung Huang from Rice University for fruitful 
	discussions and acknowledge the assistance from the S$\mu$S beamline 
	scientists. 
	This work was supported by the Schwei\-ze\-rische Na\-ti\-o\-nal\-fonds 
	zur F\"{o}r\-de\-rung der Wis\-sen\-schaft\-lich\-en For\-schung, 
	SNF (Grants No.\ 200021\_169455 and 206021\_139082).  }

\vspace{7pt}
%
\section*{References}
\bibliography{MoPB_bib}

\end{document}